\documentclass[fleqn,usenatbib]{mnras}

\usepackage{newtxtext,newtxmath}

\usepackage{siunitx}
\usepackage{rotating}
\usepackage{lscape}

\usepackage[T1]{fontenc}

\DeclareRobustCommand{\VAN}[3]{#2}
\let\VANthebibliography\thebibliography
\def\thebibliography{\DeclareRobustCommand{\VAN}[3]{##3}\VANthebibliography}

\usepackage{graphicx}	
\usepackage{amsmath}	
\usepackage[dvipsnames]{xcolor} 
\usepackage{enumitem} 

\usepackage{caption}
\usepackage{subcaption}
\usepackage{gensymb}

\title[Jet-ISM interactions and feedback in a $z\sim0.1$ quasar]{Quasar Feedback Survey: Multi-phase outflows, turbulence and evidence for feedback caused by low power radio jets inclined into the galaxy disk\thanks{Based on data obtained under ESO Proposal ID: 0103.B-0071.}}

\author[A.\,Girdhar, C.\,M.\,Harrison, V.\,Mainieri, et al.]{A.\,Girdhar,$^{1,2}$\thanks{E-mail: aishwarya.girdhar@eso.org}
C.\,M.\,Harrison,$^{3}$\thanks{E-mail: christopher.harrison@newcastle.ac.uk}
V.\,Mainieri,$^{1}$
A.\,Bittner,$^{1,2}$
T.\,Costa,$^{4}$ 
P.\,Kharb,$^{5}$ 
D.\,Mukherjee,$^{6}$  \newauthor 
F.\,Arrigoni\,Battaia,$^{4}$ 
D.\,M.\,Alexander,$^{7}$
G.\,Calistro\,Rivera,$^{1}$ 
C.\,Circosta,$^{8}$
C.\,De\,Breuck,$^{1}$ 
A.\,C.\,Edge,$^{7}$ \newauthor
E.\,P.\,Farina,$^{4}$ 
D.\,Kakkad,$^{9,10}$ 
G.\,B.\,Lansbury,$^{1}$ 
S.\,J.\,Molyneux,$^{11,1}$ 
J.\,R.\,Mullaney,$^{12}$ 
Silpa S.,$^{5}$ \newauthor 
A.\,P.\,Thomson,$^{13}$ 
S.\,R.\,Ward$^{1,2}$ 
\\
\\
$^{1}$European Southern Observatory, Karl--Schwarzschild--Stra{\ss}e 2, D-85748 Garching bei M{\"u}nchen, Germany\\
$^{2}$Ludwig-Maximilians-Universit{\"a}t, Professor-Huber-Platz 2, D-80539 M{\"u}nchen, Germany\\
$^{3}$School of Mathematics, Statistics and Physics, Newcastle University, NE1 7RU, UK\\
$^{4}$Max Planck Institut f\"ur Astrophysik, Karl--Schwarzschild--Stra{\ss}e 1, D-85748, Garching bei M\"unchen, Germany \\
$^{5}$National Centre for Radio Astrophysics - Tata Institute of Fundamental Research, Pune University Campus, Post Bag 3, Ganeshkhind, Pune 411007, India\\ 
$^{6}$Inter-University Centre for Astronomy and Astrophysics, Post Bag 4, Pune - 411007, India\\
$^{7}$Centre for Extragalactic Astronomy, Department of Physics, Durham University, South Road, Durham DH1 3LE, UK\\
$^{8}$Department of Physics \& Astronomy, University College London, Gower Street, London, WC1E 6BT, UK\\
$^{9}$European Southern Observatory, Alonso de Cordova, 3107, Vitacura Casilla 19001, Santiago, Chile\\
$^{10}$Department of Physics, University of Oxford, Denys Wilkinson Building, Keble Road, Oxford, OX1 3RH, UK\\
$^{11}$Astrophysics Research Institute, Liverpool John Moores University, 146 Brownlow Hill, Liverpool L3 5RF, UK\\ 
$^{12}$Department of Physics and Astronomy, The University of Sheffield, Hounsfield Road, Sheffield, S3 7RH, UK\\
$^{13}$Jodrell Bank Centre for Astrophysics, Department of Physics \& Astronomy, The Alan Turing Building, Upper Brook Street, Manchester M13 9PL, UK\\
}

\date{Accepted XXX. Received YYY; in original form ZZZ}

\pubyear{2021}

\begin{document}
\label{firstpage}
\pagerange{\pageref{firstpage}--\pageref{lastpage}}
\maketitle


\begin{abstract}
We present a study of a luminous, $z\,=\,0.15$, type-2 quasar ($L_{[\rm O III]}$=10$^{42.8}$ erg s$^{-1}$) from the Quasar Feedback Survey. It is classified as `radio-quiet' ( $L_{\mathrm{1.4\,GHz}}$=10$^{23.8}$\,W\,Hz$^{-1}$); however, radio imaging reveals $\sim$\,1\,kpc low-power jets (P$_{\mathrm{jet}}$= 10$^{44}$\,erg\,s$^{-1}$) inclined into the plane of the galaxy disk. We combine MUSE and ALMA observations to map stellar kinematics and ionised and molecular gas properties. The jets are seen to drive galaxy-wide bi-conical turbulent outflows, reaching $W_{80}$ = 1000\,-\,1300\,km s$^{-1}$, in the ionised phase (traced via optical emission-lines), which also have increased electron densities compared to the quiescent gas. The turbulent gas is driven perpendicular to the jet axis and is escaping along the galaxy minor axis, reaching 7.5\,kpc on both sides. Traced via CO(3--2) emission, the turbulent material in molecular gas phase is one-third as spatially extended and has 3 times lower velocity-dispersion as compared to ionised gas. The jets are seen to be strongly interacting with the interstellar medium (ISM) through enhanced ionised emission and disturbed/depleted molecular gas at the jet termini. We see further evidence for jet-induced feedback through significantly higher stellar velocity-dispersion aligned, and co-spatial with, the jet axis ($<$\,5\,$\degree$). We discuss possible negative and positive feedback scenarios arising due to the interaction of the low-power jets with the ISM in the context of recent jet-ISM interaction simulations, which qualitatively agree with our observations. We discuss how jet-induced feedback could be an important feedback mechanism even in bolometrically luminous `radio-quiet' quasars.
\end{abstract}

\begin{keywords}
galaxies: active – galaxy: evolution – galaxies: jets – quasars: general \end{keywords}



\section{Introduction}

Active galactic nuclei (AGN) are the sites of growing supermassive black holes at the centre of galaxies (\citealt{kormendyHo13}). They can produce enough energy, through accretion of matter, to exceed the binding energy of their host galaxies (\citealt{cattaneo09, bower12}). With this tremendous amount of energy available, if it can efficiently couple to the gas extending from the vicinity of the AGN to galactic scales, the AGN could cause significant impact (known as `feedback') on their host galaxies by either facilitating or suppressing the star-formation (see e.g., \citealt{binneyTabor95, ciottiOstriker97}; \citealt{alexanderhickox12, fabian12}; \citealt{harrison17}). This `AGN feedback' has become an imperative ingredient in cosmological galaxy formation simulations for them to reproduce key observables of galaxy populations and intergalactic material  (e.g., \citealt{bower06, mccarthy10, gaspari11, vogelsberger14, hirschmann14, schaye15, henriques15, taylorkobayashi15, choi18}). However, understanding how this process occurs in the real Universe, particularly in the case of the most powerful AGN (i.e., quasars; $L_{\rm bol}$\,$\geq$\,10$^{45}$\,erg\,s$^{-1}$), remains a challenge of extragalactic research.

One way that AGN are known to be capable of interacting with their host galaxy's multi-phase interstellar medium (ISM) is through driving galaxy-wide ($\geq$\,0.1\,-\,10\,kpc) energetic outflows and turbulences (e.g., \citealt{nesvadba11, liu2013, cicone18, veilleuxMaiolino20, davies20, fluetsch21, venturi2021, roy21}). Theoretically, these outflows may be driven by radiation pressure, jets or AGN-driven winds and could occur over scales from the accretion disc ($\leq$\,10$^{-2}$\,pc) through to galaxy-wide scales, where they could affect the properties of the gas in the ISM and beyond (\citealt{ishibashi18, mukherjee18b, costa18, costa20, morganti21}). However, from an observational perspective, whilst these multi-phase outflows are now seen to be common in quasar host galaxies, controversy remains over the relative role of the different possible driving mechanisms (\citealt{hwang18, wylezalekMorganti18, jarvis2019, somalwar20}). In particular in this work, we are interested in the relative importance of radio jets in driving multi-phase outflows (see e.g., \citealt{morganti05, nesvadba06, nesvadba08, mullaney2013, jarvis2019, molyneux2019, liaoGu20, venturi2021, RamosAlmeida22}).

Also from a theoretical perspective, AGN can regulate black hole growth and star-formation, (i.e., causing a `negative feedback' effect) by expelling gas through outflows, preventing gas from cooling and/or destroying star-forming ISM (e.g., \citealt{silkrees98, benson03, granato04, hopkins06, boothschaye10, fauchergiguere12, kingpounds15, costa18,costa20}). On the contrary, AGN may also induce a `positive feedback' effect due to the compression of the gas from outflows or jets, which induce localised enhancement of star-formation (e.g., \citealt{zubovas13, ishibashi13, fragile17, lacy17, wangLoeb18, gallagher19}). However, from an observational perspective, the impact of the most luminous AGN on their host galaxies' ISM and star-forming properties is a matter of ongoing debate (e.g., \citealt{husemann16, villarmartin16, maolino17, harrison17, kakkad17, cresci18, perna18, rosario18, doNascimento19, bluck20, yesuf20, scholtz21}). 

Understanding the drivers of outflows and the impact of AGN on their host galaxies requires multi-wavelength observations. Since outflows are multi-phase, different diagnostics are required to study their properties and impact on their host galaxy. For example, spatially-resolved observations in a small number of systems observed with both integral-field spectroscopy (IFS) and interferometers have shown that the ionised gas and molecular gas kinematics could be completely decoupled (\citealt{shihRupke10, sun14}). This further highlights the need to study multiple gas phases at comparable resolutions to fully understand the impact of the feedback. In this work, we will trace warm ($\sim$10$^{4}$\,K) ionised gas kinematics using optical emission-lines, mostly [O~{\sc iii}], which has long been used as a diagnostic to search for non-gravitational gas kinematics in AGN host galaxies (e.g., \citealt{weedman70,stockton76,heckman81,vrtilek85,borosonGreen92,veilleux95,harrison14,jarvis2019, scholtz21}). For tracing cold (i.e., tens of Kelvin) molecular gas, following many other works, we will make use of the CO emission-line, specifically the {}$^{12}$CO\,J\,=\,3\,--\,2 transition as a tracer (e.g., \citealt{combes14, sun14, cicone21,circosta21}).

This work is part of the Quasar Feedback Survey (QFeedS; Figure~\ref{fig: mullaneyplot};  \citealt{jarvis21}). This survey aims to address the challenges outlined above by studying the spatially-resolved multi-wavelength properties of 42 relatively low redshift ($z<0.2$) quasar host galaxies. Using this redshift range allows us to obtain sensitive, high spatial resolution observations (typically $\lesssim$1\,kpc), whilst also yielding a reasonable sample of powerful quasars ($L_{\rm bol}\gtrsim10^{45}$\,erg\,s$^{-1}$) representative of $L_{\star}$ at the peak cosmic epoch of growth, where quasar feedback is expected to dominate (\citealt{kormendyHo13}). Several pilot studies have been conducted on a subset of the QFeedS targets \citep[][]{harrison14, harrison15, lansbury18, jarvis2019, jarvis2020}, whilst an overview of the full sample can be found in \cite{jarvis21}.  Spatially-resolved radio observations form the basis of QFeedS, which  provides important insights into the prevalence and properties of radio jets in what is a representative and predominantly `radio-quiet' quasar sample \citep{jarvis2019,jarvis21}. Here we define radio-quiet using the criterion of \citealt{xu1999} that compares the radio to [O~{\sc iii}] emission-line luminosity. For a detailed discussion in the context of the QFeedS sample see \citealt{jarvis21}.

In this work we study one target from QFeedS for which we combine published radio observations with new observations from the integral field spectrograph Multi Unit Spectroscopic Explorer (MUSE; \citealt{Bacon10,bacon2017}) and from the Atacama Large Millimeter/submillimeter Array (ALMA). The former enables us to map the stellar kinematics and ionised gas properties and the latter enables us to map the molecular gas properties. In Section \ref{sec: data}, we introduce the target, observations and data reduction. In Section \ref{sec: jetprop} we estimate radio jet properties from existing data. In Section \ref{sec: methods} we describe the approaches employed to fit the spectra and map the stellar kinematics and gas properties. In Section \ref{sec: results} and Section \ref{sec: discussions}, we present our results and discussion, respectively. Finally, in Section \ref{sec: conclusions}, we summarise our conclusions. 

We have adopted the cosmological parameters to be $H_0$ = 70 km s$^{-1}$ Mpc$^{-1}$, $\Omega_M$ = 0.3 and $\Omega_{\Lambda}$ = 0.7, throughout. In this cosmology, 1\,arcsec corresponds to 2.615 kpc for the redshift of $z=$\,0.15 (i.e., the redshift of the galaxy studied here).  We define the radio spectral index, $\alpha$, using $S_{\nu}\propto \nu^{\alpha}$. 


\section{Target, Observations and Data Reduction}\label{sec: data}

In Section~\ref{sec: qfeeds} we describe the target investigated in this study in the context of the parent Quasar Feedback Survey and summarise its basic radio properties. We describe the details of the observations and data reduction for the MUSE and ALMA data in Section~\ref{sec: muse} and ~\ref{sec: datared_alma}, respectively. A summary of the data used throughout this work is presented in Figure~\ref{fig: J1316}, with the central panel showing an optical continuum image from the MUSE data ($\sim$62$\times$62\,kpc); the right panel showing a closer look ($\sim$10$\times$10\,kpc) of the pseudo narrow-band [O~{\sc iii}] image (from MUSE) with CO (3--2) emission (from ALMA) overlaid as contours in purple; and the left panel with a zoom-in ($\sim$26$\times$26\,kpc) of the continuum image, focused on the central regions of the galaxy. A description of how these images were created is described in the following sub-sections and Section~\ref{sec: fig2}.

\subsection{J1316+1753 in the context of the Quasar Feedback Survey}\label{sec: qfeeds}

\begin{figure}
	\includegraphics[width=\columnwidth]{ 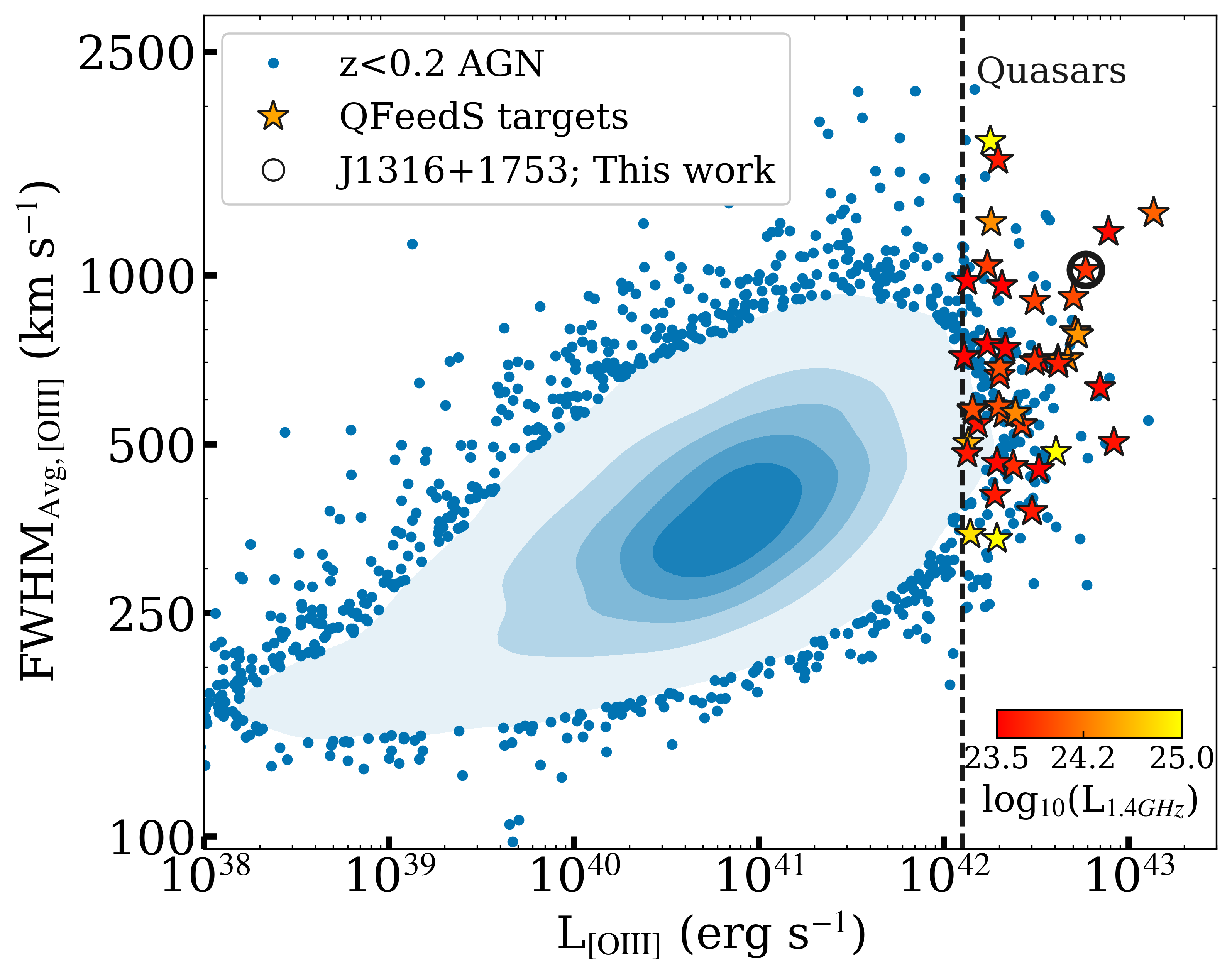}
    \caption{[O~{\sc iii}] FWHM versus $L_{\rm [O~III]}$ for $z\,<\,0.2$, spectroscopically-selected AGN from \citealt{mullaney2013}, represented as blue data points and density contours. Stars represent the 42 QFeedS targets, with $L_{\rm [O~III]}>10^{42.1}$\,erg\,s$^{-1}$ (dashed line), from which the quasar of this study was selected (highlighted with a black circle). The QFeedS targets are colour-coded by their radio luminosities ($L_{\rm 1.4GHz}$). The quasar in this study has a modest, representative radio luminosity ($L_{\rm 1.4GHz}$=10$^{23.8}$ W\,Hz$^{-1}$) and it is classified as `radio-quiet' using traditional criteria (see Section \ref{sec: qfeeds}).}
    \label{fig: mullaneyplot}
\end{figure}

\begin{figure*}
	\includegraphics[width=\textwidth]{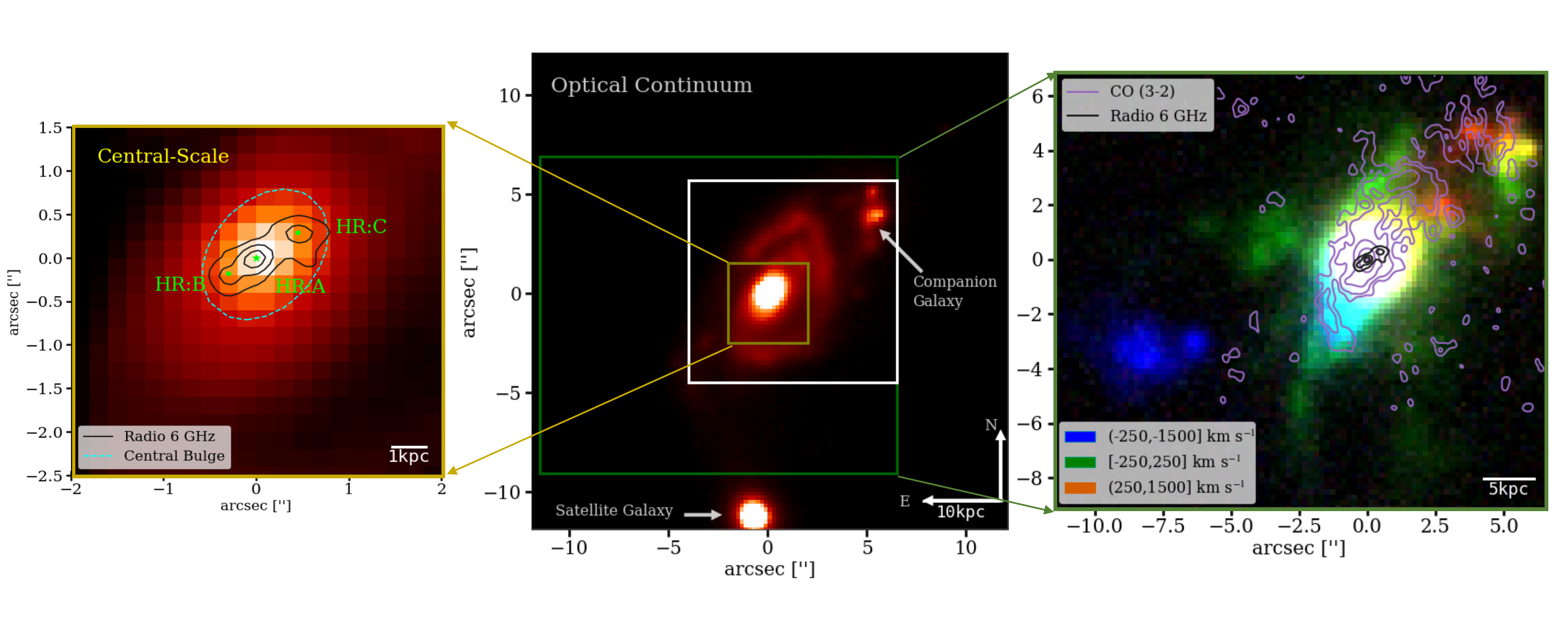}	
	\caption{A multi-wavelength and multi-scale view of our target galaxy. {\em Centre:} An optical continuum image from the MUSE data showing the local, 24$\times$24\,arcsec ($\sim$62$\times$62\,kpc), environment of J1316+1753. The primary target galaxy is centered at a position (0,0) with two clear spiral arms. To the north-west is an associated companion galaxy and to the south one passive satellite galaxy can be seen. {\em Right:} A closer look (18$\times$16\,arcsec; $\sim$\,47$\times$42\,kpc) showing the pseudo narrow-band [O~{\sc iii}] image (from MUSE) with CO (3--2) emission (from ALMA) overlaid as contours in purple (with levels [16,8,4,2,1]\,RMS$_{\mathrm{CO(3-2)}}$). The three colours are obtained by dividing the [O~{\sc iii}] emission-line profile in three velocity slices as indicated in the legend. {\em Left:} A zoom-in of the continuum image in the central regions of the galaxy (4$\times$4\,arcsec;  $\sim$\,10$\times$10\,kpc). The black contours are the VLA 6\,GHz radio emission, at the levels [128,64,32]\,RMS$_{\mathrm{radio}}$. In light green we show three dots to highlight the positions of the radio core (HR:A) and the two radio jet hotspots (HR:B and HR:C), as defined in \citealt{jarvis2019}. The central panel also shows the `galaxy-scale' (white inset; 10$\times$10\,arcsec; $\sim$\,26$\times$26\,kpc) within which the kinematic analysis is done. North is up and East is left in all panels and for all plots throughout this paper.}
    \label{fig: J1316}
\end{figure*}

This paper focuses on a single target, J1316+1753, with an optical right ascension (RA) of 13$^{\rm{h}}$16$^{\rm{m}}$42$^{\rm{s}}$.90 and declination (DEC) of $+$17$^{\degree}$53'32''.5 (J2000, ICRS). It is selected from QFeedS and this target illustrates the rich information gained on outflows and AGN feedback from our high-resolution multi-wavelength QFeedS dataset, comprising of radio, optical and sub-mm data. QFeedS was originally constructed using the emission-line selected AGN from \cite{mullaney2013}. Figure \ref{fig: mullaneyplot} presents the 16,680 $z<0.2$ AGN from \cite{mullaney2013} in the [O~{\sc iii}] full-width-half-maxima (FWHM$_{\rm Avg}$) versus [O~{\sc iii}] luminosity parameter space. The QFeedS targets are indicated with stars in the figure. We note that the FWHM$_{\rm Avg}$ values plotted in Figure \ref{fig: mullaneyplot} are the flux-weighted average of the FWHMs of the two emission-line components fitted by \cite{mullaney2013}. QFeedS targets were selected to have quasar-level [O~{\sc iii}] luminosities (i.e., $L_{\rm [OIII]}$ > 10$^{42.1}$ erg s$^{-1}$; \citealt{jarvis21}). As described in \cite{jarvis21}, a radio luminosity selection of $L_{\rm 1.4GHz} > 10^{23.45}$\,W Hz$^{-1}$ is also applied to obtain the final QFeedS sample of 42 targets. This radio cut was set to be above the NVSS detection limit (\citealt{mullaney2013, molyneux2019}). Despite their moderate radio luminosities, the QFeedS sample still consists of 88\,\% `radio-quiet' sources, based on the criteria of \cite{xu1999} (see \citealt{jarvis21}), which is consistent with the ‘radio-quiet’ fraction of the overall quasar population (i.e. $\sim$90\,$\%$; \citealt{zakamska04}). Furthermore, the QFeedS targets cover a representative range of emission-line widths of the parent population.

The target of this study, J1316+1753, is highlighted with a circle in Figure~\ref{fig: mullaneyplot}. A comprehensive summary of the target's properties and its previous studies can be found in the supplementary material of \cite{jarvis21}. However, briefly, J1316+1753 has a type-2 spectra; it has an optical emission-line redshift of $z\,=\,0.150$\, (see \citealt{jarvis2019}); it is one of the most optically luminous sources in the QFeedS sample (with $L_{\rm [O III]}$ = 10$^{42.8}$\,erg\,s$^{-1}$); and it exhibits a broad [O~{\sc iii}] emission-line profile (FWHM$_{\rm Avg}=1022$\,km\,s$^{-1}$). The one-dimensional [O~{\sc iii}] emission-line profile consists of two bright narrow emission-line components, in addition to a strong broad emission-line component (see \citealt{jarvis2019}) and we investigate this emission using our spatially-resolved MUSE data in Section \ref{sec: methods}. Based on the spectral energy distribution (SED) fitting from the \textit{UV} to \textit{FIR}, J1316+1753 is a star-forming galaxy with star formation rate $=\,30\pm10\,$M$_{\odot}$\,yr$^{-1}$, and stellar mass $=\,10^{11}$\,M$_{\odot}$ (see \citealt{jarvis2019} for these estimates).

Importantly, J1316+1753 has a modest radio luminosity of L$_{\rm 1.4GHz}$= 10$^{23.8}$ W\,Hz$^{-1}$ and, although it is confirmed to host a radio AGN, it is classified as `radio-quiet', as per the traditional criteria \cite[][]{xu1999,bestheckman12,jarvis21}. Compared to some of the other QFeedS sources, J1316$+$1753 has relatively simple radio morphology \citep[][]{jarvis2019,jarvis2020}. The 6\,GHz radio image of this target from the Very Large Array (VLA) data was produced by \cite{jarvis2019} (with a synthesised beam of 0.33$\times$0.28\,arcsec; root-mean-square (RMS) noise of 8\,$\mu$Jy\,beam$^{-1}$) and the morphology reveals three components, which are most likely attributed to a radio core and two hot spots associated with radio jets. These were labelled as HR:A, HR:B and, HR:C in \cite{jarvis2019} respectively, and we continue with this nomenclature in this work. The hot spots HR:B and HR:C are located at a projected distances of 0.91\,kpc and 1.40\,kpc from the central component, respectively. The central radio component, HR:A, has a flatter 1.5--7\,GHz spectral index of $\alpha\sim-0.3$, compared to the two hot spots, which have $\alpha\sim-1$ and $\alpha\sim-1.2$ (for HR:B and HR:C, respectively; \citealt{jarvis2019}). Overall, the properties of J1316+1753 provide a cogent case to investigate the impact of low luminosity and compact radio jets on the multi-phase ISM in a `radio-quiet' quasar host galaxy. 

\subsection{Observation and reduction of the MUSE data}\label{sec: muse}

J1316+1753 was observed with the Multi Unit Spectroscopic Explorer \citep[MUSE;][]{Bacon10} on ESO's Very Large Telescope (VLT) as part of proposal ID: 0103.B-0071 (PI: Harrison). MUSE is an optical and near-IR integral-field spectrograph at the VLT, housed at the European Southern Observatory's Paranal Observatory. The observations were carried out in wide field mode, with adaptive optics, in a single observing block on the night of 2019 April 29-30. The observations consisted of 4 science exposures (of 550\,s each) and 1 sky exposure (of 100\,s), rotated by 90$\degree$ and dithered within a $\sim$4\,arcsec box. The spectra ranges from 4750 to 9350 \AA\ with a spaxel size of about 1.25 \AA\ . The field-of-view is about 1$\times$1 arcmin, with a pixel size of 0.2\,arcsec.\

The raw data were reduced using the standard ESO MUSE pipeline version 2.6 \citep{weilbacher14, weilbacher20}. The flow of the pipeline includes bias-subtraction, correction for flat-field, illumination and geometry, and calibration in wavelength and flux. The sky emission was estimated using the sky exposure and subtracted from the science data. The datacubes were resampled onto a $0.2\arcsec \times 0.2\arcsec \times 1.25$\,\AA\,grid. For the flux scaling, the relative offsets between the four exposures were computed using an average of 12 sources in the white-light images of each datacube. The final datacube was obtained by average-combining the four exposures corrected for their offset. 

To estimate the seeing point spread function (PSF) for our cube, we used two point-like sources in our field of view. We estimated the PSF to be $\sim$\,0.75\,arcsec, in the wavelength range around the [O~{\sc iii}]$_{5007}$ emission, using the \texttt{MoffatModel2} function from the \texttt{mpdaf} package (\citealt{mpdaf}). We corrected for the wavelength-dependent line-spread function (LSF), introduced by MUSE, during the line-profile fitting procedure, as described in Section \ref{sec: emline}.

\subsection{Observation and reduction of the ALMA data}\label{sec: datared_alma}

J1316+1753 was observed with ALMA in Band 7 in $3\times1$\,hour epochs between 2019 April 29\,-\,2019 May 03 under ALMA project 2018.1.01767.S (PI: A.\,P.\ Thomson). Our chosen correlator setup comprises of three spectral windows, with one having a central frequency $\nu_{\rm obs}=345.795990$\,GHz, corresponding to the redshifted CO(3-2) line at $z_{\rm opt}=0.15$ \citep{jarvis2019}, and the others overlapping this spectral window with central frequencies of 344.7\,GHz and 347.0\,GHz used for baseline subtraction. Each of the three spectral windows has a bandwidth of 1875\,MHz (1870\,km\,s$^{-1}$) and a spectral resolution of 31.25\,MHz (31.17\,km\,s$^{-1}$), recorded in dual polarization (XX,YY). In total, our three overlapping spectral windows provide 4.17\,GHz ($\sim 3600$\,km\,s$^{-1}$) of spectral coverage centred on the CO(3-2) line. Our science goals of mapping the galaxy-scale molecular gas kinematics and looking for jet-ISM interactions in the region of the radio jet, lead to using the C43-4 array, providing an angular resolution of $\theta_{\rm res}\,\sim\,0.5$\,arcsec and a largest angular scale $\theta_{\rm LAS}\sim4$\,arcsec (corresponding to linear scales of $\sim1.3$\,kpc and $\sim 11$\,kpc, respectively, at the source redshift $z=0.15$). 

We obtained the raw data from the ALMA science archive and pipelined the data in {\sc CASA v5.4.0-68} using the reduction script supplied by ALMA user support. Following the successful completion of the ALMA CASA pipeline, the phases and amplitudes were visually inspected as functions of frequency and time using the {\sc casa} tool {\sc plotms}, during which a small percentage ($\lesssim 1\%$) of outliers were flagged from the calibrated data.

The reduced $uv$ data were imaged in {\sc casa} {\sc tclean}. A spectral line datacube was created, using natural weighting ({\sc robust}=2.0), on-the-fly primary beam correction ({\sc pbcor}=True) and multi-scale clean at a pixel scale of 0.05\,arcsec\,pix$^{-1}$, across an image of $\sim51$\,arcsec, (or 135\,kpc) on a side. The resulting naturally-weighted images have a synthesized beam which well-resembles a two-dimensional Gaussian of width $0.54''\times 0.45''$ at a position angle of $8^\circ$, with an rms sensitivity of $\sigma_{\rm rms}=347\,\mu$Jy\,beam$^{-1}$\,channel$^{-1}$.

\subsection{Alignment of the MUSE, ALMA and VLA data}\label{sec: astrometry}
To perform a multi-wavelength analysis we need to be sure that all the data are registered on the same astrometric solution. To do this, we followed the same procedure as in  \citealt{jarvis2019}, who demonstrated that the astrometric solutions of Sloan Digital Sky Survey (SDSS) images were accurate enough to align datacubes to VLA data, for 9 targets from QFeedS including J1316+1753, with a median accuracy of 0.13\,arcsec (i.e., $\sim$0.3\,kpc). This is sufficient for the multi-wavelength comparison that we perform in this work (Section \ref{sec: methods}). Briefly, we created pseudo broad-band images from the MUSE IFS cubes over the same wavelength range as the \textit{g}-band images from SDSS. The peak emission in the pseudo-broad band image was identified to be located 0.6706\,arcsec from the peak radio emission and 0.6709\,arcsec from the peak CO(3--2) emission (see Figure~\ref{fig: J1316}; Section \ref{sec: fig2}). Then we used the \texttt{find\_peaks} function of the \texttt{photutils} (\citealt{photutils}) python package to identify the pixel associated to the peak emission in the MUSE pixel and anchored the coordinate values of this pixel to match the associated position from the image peak in SDSS. This resulted in a global astrometric shift of $\sim$\,2.5\,arcsec to the MUSE cube.

Since, after this shift, the peak continuum and the radio core were well aligned (i.e., within $\sim$\,0.1\,arcsec) this provides confidence in the alignment process, although we note that a priori these different emissions may not by physically aligned. Ultimately we chose the HR:A position of the radio core (introduced in Section \ref{sec: qfeeds}; RA of 13$^{\rm{h}}$16$^{\rm{m}}$42$^{\rm{s}}$.8 and DEC of 17$^{\degree}$53'32''.3) as our `nucleus' position, indicated with a (0,0) in all relevant figures. All distances are measured with respect to this position as the reference point.

\section{Estimated jet properties}\label{sec: jetprop}

In this section, we estimate the properties of the radio jets obtained from empirical relations and existing radio data for the target. These estimated properties are used to help interpret our new observations in terms of the effects of the jets on the quasar host galaxy in Section \ref{sec: energetics}.  

\subsection{Jet Kinetic Power ($P_{\mathrm{jet}}$)}
To estimate the kinetic jet power ($P_{\mathrm{jet}}$), we use the empirical relation of \citealt{merloniheinz2007}, obtained from a sample of low-luminosity radio galaxies, as below: 
\begin{equation}\label{eq: pjet}
   \log P_{\mathrm{jet}} = (0.81\,\pm\,0.11)\,\times\,\log L_{5\,\mathrm{GHz}}\,+\,11.9_{-4.4}^{+4.1}
\end{equation}

From the VLA data in \citealt{jarvis2019}, the 5.2\,GHz flux density of the radio core (HR:A) is 1.1\,mJy (spectral index $\alpha$\,=\,-\,0.3). This corresponds to 5\,GHz luminosity of  6.5\,$\times$\,10\,$^{39}$\,erg\,s$^{-1}$ for the radio core. Using the relation above we estimate a jet power of 7.8\,$\times$\,10\,$^{43}$\,erg\,s$^{-1}$. However, since the empirical relation is derived from lower resolution data than ours, we also calculate the jet power using the total 5\,GHz flux density (i.e., over HR:A, HR:B and HR:C), which is 2.7\,mJy (spectral index $\alpha$\,=\,-\,0.85; \citealt{jarvis2019}). Using this flux density, we estimate a jet power that is only higher by a factor of 2.5 (i.e., 1.7\,$\times$\,10\,$^{44}$\,erg\,s$^{-1}$). Hence, for our order-of-magnitude estimate, we assume the jet power to be, $P_{\mathrm{jet}}$\,=\,10\,$^{44}$\,erg\,s$^{-1}$ during our discussion in Section~\ref{sec: energetics}.

\subsection{Jet inclination angle\,($\theta$)}\label{sec: jetangle}
Since AGN jets experience bulk relativistic motions (\citealt{blandfordkonigl79, ghisellini93}), the apparent brightness of the different AGN sub-components can be used as an estimator of their inclination angle with respect to line of sight (los). The radio core prominence parameter is a measure of the core-to-extended radio flux density [\(R_{\rm{C}} = S_{\rm{core}}/S_{\rm{ext}}\,=\,S_{\rm{core}}/(S_{\rm{tot}}-S_{\rm{core}})\)], and is a known statistical indicator of inclination angle, with respect to los, as discussed in detail in \citealt{orrbrowne82} (also see \citealt{kapahisaikia82,kharbshastri2004}). We used the flux density estimates from the 5.2 GHz VLA data from \citealt{jarvis2019}, which estimates the $S_{\rm{core}}$\,=\,1.1\,mJy and $S_{\rm{tot}}$\,=\,2.7\,mJy.  Using these values we estimated the $R_{\rm{C}}$ to be = 0.7, which gives \(log_{10}(R_{\rm{C}})=-0.16\). The $log\,R_{\rm{C}}$ values are seen to vary from -4 to +4 for a sample of randomly oriented radio galaxies and blazars, corresponding to inclination angles from 90\,\degree\, to 0\,\degree\, (see \citealt{kharbshastri2004}). Therefore, the estimated \(logR_{\rm{C}}\,\sim\,0\), value for J1316+1753 corresponds to an inclination angle of $\sim$\,45\degree\, for the jet in this source, with respect to our line of sight. Hence, based on these statistical arguments, a projected inclination of $\sim$\,45\,\degree\ with respect to our los, is assumed for the jets in the rest of the paper. Since we estimated our galaxy's position angle to be 46\,\degree\,(see Section~\ref{sec: angle_size}), we conclude that the jets are seen to be in the plane of the galaxy disk.

\subsection{Jet Speed\,($\beta$)}\label{sec: jetspeed}
The relativistic doppler beaming effects in jets enable us to estimate which is the approaching and which is the receding jet, and to also estimate a jet speed (\citealt{ghisellini93,urrypadovani95}). Using the 5.2 GHz VLA image from \citealt{jarvis2019}, we measured the surface brightness (SB\,[$\micro$\,Jy\,/\,beam]) values at equal distances from the peak surface brightness point (i.e., the radio core, HR:A) on both sides along the jet axis. We define the ratio of these surface brightness values as \(J = SB_{\rm{HR:B}}/SB_{\rm{HR:C}}\), and repeated these measurements at 7 different distances ranging from 0.07-1.2\,arcsec. A range of $J$ estimates was obtained from 1.1 to 2.6. On average, the south-western jet (HR:B) was measured with $\sim$\,2 times higher surface brightness than the north-eastern jet (HR:C). Hence it was concluded that HR:B is the approaching side of the jet.

Furthermore, the ratio, J was used to estimate the jet speed, $V_{\rm{jet}}=\beta\,c$, using equation A10 from \citealt{urrypadovani95} as follows:
\begin{equation}
\beta\,\cos\,\theta = \dfrac{\mathrm{J}^{1/p} - 1}{\mathrm{J}^{1/p} + 1}
\end{equation}
We assumed the value of \(p = 3.0\) (as defined for a continuous, non-blobby jet, with $\alpha$=$-$1; \citealt{urrypadovani95})\footnote{We note that \citealt{urrypadovani95} use the opposite sign convention for spectral index ($\alpha$) in their study compared to this analysis.}. Using our estimate of the jet inclination angle ($\theta=$\,\,45\,\degree; see Section \ref{sec: jetangle}), we estimate the jet speed to be in the range 0.04\,$c$\,--\,0.2\,$c$. This seems to be a reasonable value, since for a typical low-luminosity AGN, jet speeds would be expected to be around 0.15\,$c$ (\citealt{ulvestad99}). A limitation of the jet speed and jet inclination estimates is that this relation is defined for radio-loud AGNs, in contrast to our radio-quiet source. However, we mainly use these values for a qualitative comparison to simulations and order-of-magnitude estimates in Section~\ref{sec: energetics}.


\begin{figure*}
    \centering
    \hspace{0.3cm}
    \begin{subfigure}{1\columnwidth}
        \centering
        \includegraphics[height=6cm]{ 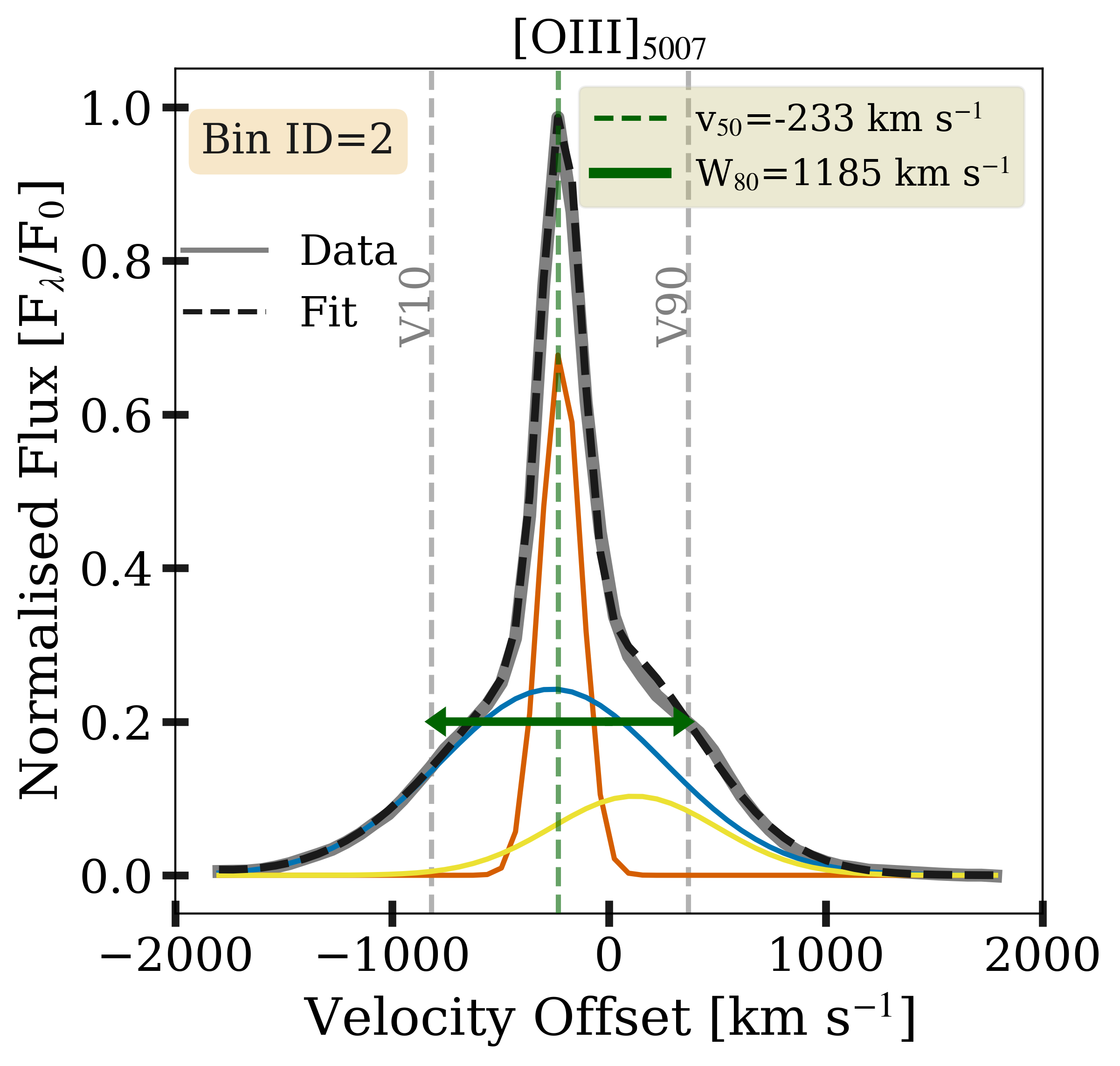}
    \end{subfigure}%
    ~ 
    \hspace{-1.2cm}
    \begin{subfigure}{1\columnwidth}
        \centering
        \includegraphics[height=6cm]{ 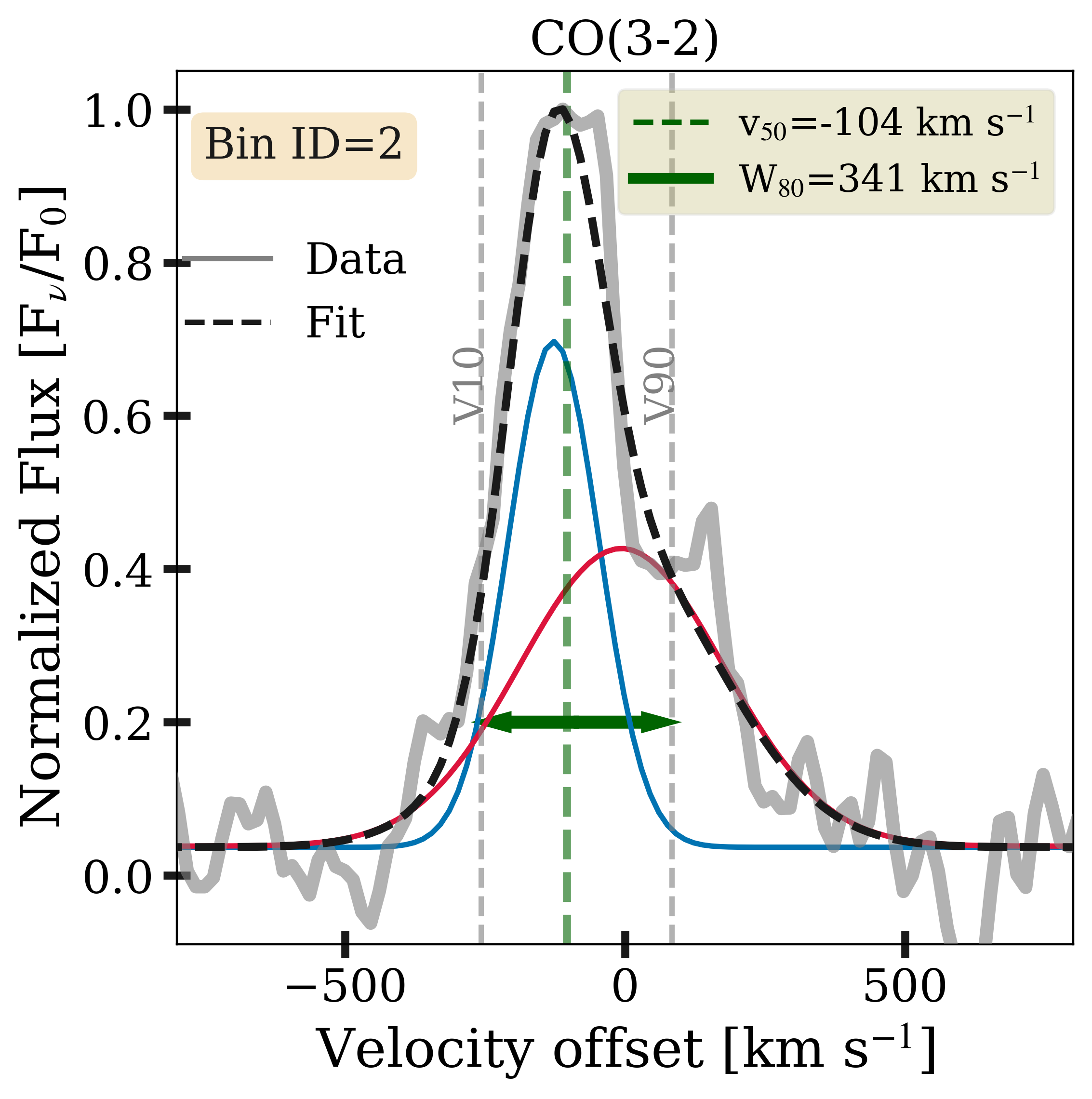}
    \end{subfigure}
    ~ 
    \begin{subfigure}{2\columnwidth}
        \centering
        \includegraphics[height=5.5cm]{ 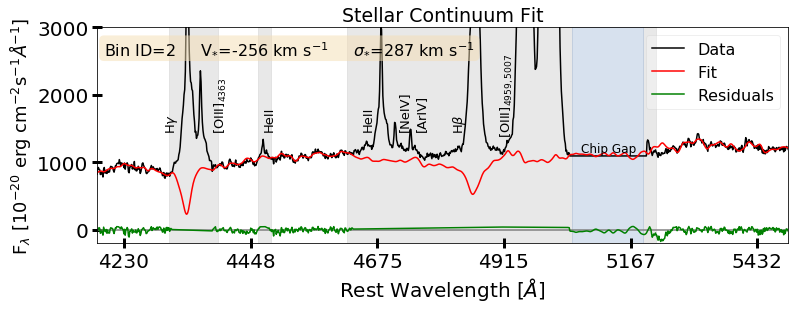}
    \end{subfigure}
    \caption{Examples of our emission-line and stellar continuum fits to the MUSE and ALMA data. {\em Upper-left:} [O~{\sc iii}] emission-line profile (grey curve) and our best fit (dashed black curve), as well as the three individual Gaussian components used to construct the total fit (yellow, blue and orange curves).  {\em Upper-right:} CO(3--2) emission-line profile (grey curve), with the best fit (dashed black curve) and the two Gaussian components that make up the total fit (blue and red curves). In both upper panels we indicate the non-parametric values obtained from our analyses: $V_{50}$ (dashed green line) to trace the bulk velocity and $W_{80}$ (green arrows) to trace the velocity-dispersion, which is the velocity width between $V_{10}$ and $V_{90}$ (vertical dashed grey lines). Both emission-line profiles are extracted from Voronoi Bin no. 2, as defined in Section \ref{sec: gist}. {\em Lower panel:} Spectra (black curve), presented across the wavelength range used to obtain the stellar continuum fits (red curve) for estimating the stellar kinematics (see Section \ref{sec: gist}). Wavelength regions excluded from the fits (due to strong emission-lines and the instrumental chip gap) are highlighted with shaded regions. The fit residuals are shown as green curves.}
    \label{fig: fits}
\end{figure*}

\section{Analysis}\label{sec: methods}

In this section, we describe our analysis methods to: (1) produce the continuum and emission-line images shown in Figure~\ref{fig: J1316} (Section~\ref{sec: fig2}); (2) estimate the structural parameters of the galaxy (Section~\ref{sec: angle_size}); (3) map the stellar kinematics using the MUSE data (Section \ref{sec: gist}); (4) model the optical and CO(3--2) emission-line profiles from the MUSE and ALMA data, respectively (Section \ref{sec: emline}); (5) estimate the non-parametric properties of the emission-lines and produce their maps (Section \ref{sec: nonparam}). As illustrated in Figure~\ref{fig: J1316}, we perform the analyses on two spatial scales, firstly over the whole  `galaxy-scale' (for which we map the stellar and gas properties using bins following Voronoi tesellation, see Section \ref{sec: gist}) and, secondly, focusing on the inner `central-scale' region around the jets (for which we map the gas properties only).

\subsection{Environment and central-regions of J1316+1753}\label{sec: fig2}

We present pseudo-broadband and narrow-band (emission-line) images of J1316+1753 in Figure \ref{fig: J1316}. We note that for all images in this paper, North is up and East is to the left, as indicated in the central panel of this figure. The central panel shows the stellar continuum image of the galaxy inside a 24$\times$24\,arcsec\ (62$\times$62\,kpc) region. To make the image, we median-collapsed the data cube (masking all the emission-lines) across the same wavelength range as that used for the stellar continuum fitting (see Section \ref{sec: gist} for a detailed explanation and Figure \ref{fig: fits} for a representation). The figure clearly reveals the galaxy morphology, with a central bulge and two prominent spiral arms extending over $\sim$ 21\,kpc. Within this image we identify, in addition to our main target, one passive satellite galaxy to the south (without any emission-lines). However, based on the regular morphology and kinematics (discussed in Section \ref{sec: res_gal}), we do not see any clear disturbances in the galaxy dynamics caused by it that could affect our conclusions and hence they are not discussed henceforth in this work. We make particular note of the bright and clumpy continuum structure located $\sim$\,17\,kpc to the north-west of the primary galaxy nucleus. For identification purposes only, we label this as the `companion galaxy'. We note that this is identified in emission lines at the same redshift as our target galaxy, meaning that it is also physically associated. The kinematic anaysis in Section \ref{sec: res_gal} is done within the region indicated through the white box. We refer to it as the `galaxy-scale' region (10$\times$10\,arcsec\ or 26$\times$26\,kpc).

The right panel of Figure~\ref{fig: J1316} covers a spatial region of 18$\times$16\,arcsec (i.e., 47$\times$42\,kpc), chosen to show the extent of the [O~{\sc iii}] gas we identified in and around the main galaxy. The three-colour image shown is comprised of three pseudo narrow-band [O~{\sc iii}] images, collapsed over the velocity slices indicated in the legend. The overlaid contours in purple represent the CO(3--2) image, created by collapsing the ALMA cube over a frequency range of $\pm$700\,km\,s$^{-1}$. The black overlaid contours are from the 6 GHz VLA image (see Section \ref{sec: qfeeds}; \citealt{jarvis2019}).

In the left-panel of Figure~\ref{fig: J1316} we show a zoom-in of the continuum image, as indicated by the yellow box in the main panel, extending over $\sim$\,4\,arcsec\,(i.e., 10\,kpc). We use this region as the `central-scale' to perform our spaxel-by-spaxel kinematic analyses in Section \ref{sec: res_nuc}. The three light-green color dots indicate the position of the radio core (HR:A) and the two jet hotspots (HR:B and HR:C; discussed in detail in Section \ref{sec: qfeeds}). The cyan color ellipse shows an isophote fit to the stellar bulge profile, and is discussed later in Section \ref{sec: angle_size}.

\subsection{Position angle, inclination angle and sizes}\label{sec: angle_size}

To aid the discussion on how the jets and gas interact, it is useful to constrain some basic properties of the host galaxy stellar emission, specifically the size, major axis position angle and inclination angle. We used the high spatial resolution continuum image (Figure \ref{fig: J1316}; middle panel) to make measurements of half-light radii ($R_{1/2}$), morphological axes (PA) and inclination angle ($\theta$). We fitted a two-dimensional Gaussian model to our galaxy's continuum image to obtain a morphological position angle $=$\,152\,$\pm$\,2$\degree$, an axis ratios (b/a) $=$\,0.71\,$\pm$\,0.04, and $R_{1/2}$\,=\,6.10\,$\pm$\,0.12\,kpc. We note that all quoted position angles are defined as going East of North. The half-light radius estimate was further verified by fitting an exponential Sersic profile to the surface brightness of the galaxy disc. Furthermore, following the works of \cite{tullyFisher77,weijmans14}, we used the axis ratios ($b/a$) to estimate the inclination angle ($\theta$) using the following relation: 
\begin{equation}
\cos^2 \theta = \dfrac{(b/a)^2 - q_0^2}{1 - q_0^2}
\end{equation}

where q$_0$ is the intrinsic axial ratio of an edge-on galaxy (\citealt{tullyFisher77}). We assumed $q_0$\,=\,0.2, which is appropriate for thick disks, however, for our axis ratio values, a change in $q_0$ by a factor of 2 would result in a $<$\,9\,\% change in the inclination angle ($\theta$). Using the above relation, we estimated a galaxy inclination angle of 46\,$\pm$\,2$\degree$. We would like to note that the reported uncertainties are purely from the fitting function, and do not take into account the intrinsic uncertainties arising due to the choice of a simple model.

In Section \ref{sec: positivefeedback}, we are interested in the relative position angle between the radio jets and the observed stellar bulge. Therefore, we followed a similar approach for the stellar bulge (Figure \ref{fig: J1316}; left-panel) and fitted a 2D Gaussian (cyan ellipse) to obtain a position angle of 139\,$\pm$\,4$\degree$. 

\subsection{Stellar kinematics from MUSE}\label{sec: gist}

The gas kinematics in centres of galaxies can be a good tracer of the AGN-driven outflows and the consequent jet-ISM interactions. However, the effect of the underlying stellar kinematics needs to be quantified to de-couple gravitational-motions from gas dynamics arising due to other processes. We use the MUSE data to map the stellar kinematics over the full extent of the stellar emission, i.e., the `galaxy-scale' indicated by the white box in Figure~\ref{fig: J1316}.

We fitted the stellar continuum using the modular analysis framework of \texttt{GIST} pipeline\footnote{\url{http://ascl.net/1907.025}} \citep[Galaxy IFU Spectroscopy Tool;][]{bittner2019}. As a brief overview, the \texttt{GIST} pipeline works directly on the science-ready MUSE datacubes, performs a Voronoi tesellation (\citealt{capcop2013}), fits spectral templates to the observed spectra and outputs a best-fit estimate of stellar velocity and stellar velocity-dispersion for each Voronoi bin. A complete overview of the pipeline is presented in \cite{bittner2019}. 
 
The spectra were de-redshifted to the rest-frame using an initial estimate of the systemic redshift ($z=0.15$). The spaxels below an isophote level with an average signal-to-noise-ratio (SNR) of 5 per spectral bin, as measured within the rest-frame wavelength range of 4452 to 4630 {\AA}\, (range devoid of any bright emission-lines), were excluded from the fits. Then, the remaining galaxy spaxels were spatially binned to achieve a minimum SNR using the adaptive Voronoi tesellation routine of \cite{capcop2013}. We set the minimum SNR per Voronoi bin to be 30, which resulted in stable measurements of the stellar kinematics, but still provided a good spatial sampling (see Section~\ref{sec: res_gal}). As a result of this procedure, the galaxy was divided in 41 Voronoi bins over the galaxy-scale frame. We show the shape and location of these bins in the Appendix, Figure~\ref{fig: voronoi_map}. Of these, 38 bins (Bin No. 0-37) were contiguous and represented the main galaxy under investigation, while 3 bins (Bin No. 38, 39, 40) were associated with the minor companion galaxy (as discussed in Section \ref{sec: galaxy_overview}, and indicated in Figure \ref{fig: J1316}). 

The actual wavelength range used to fit the stellar continuum was restricted to 4185--5500 {\AA} in the rest-frame. This wavelength range covers sufficient emission-line free regions of prominent stellar absorption features, whilst avoiding the redder wavelength range of MUSE that is more strongly affected by sky line residuals \citep[also see e.g.,][]{falconbarroso2006,bittner2019}. 

The stellar continuum fitting with \texttt{GIST} requires the preparation of a spectral template library. We required a spectral template library that is compatible with sufficient wavelength range and the spectral resolution for our MUSE data. Following these requirements, we selected the second data release (DR2) of the X-Shooter Spectral Library \citep[\texttt{XSL;}][]{arentsen2019,gonneau2020}. The \texttt{XSL} is a stellar spectral library covering the wavelength range from 3000-25000 {\AA}, at a resolution of about R $\sim$ 10000. We used the ultra-violet to blue (UVB) set of the DR2 of \texttt{XSL} that covers 813 observations of 666 stars. The parametrisation from \cite{bacon2017} was adopted as the MUSE line-spread function to broaden all the template spectra to the resolution of our observed spectra before any fits were conducted. This process corrects the velocity-dispersion measurements for the spectral resolution of MUSE.

The stellar kinematics were derived by performing a run of the \texttt{pPXF} routine \citep{capem2004, cappellari17}. To account for small differences between the MUSE spectra and the templates, we included an additive Legendre polynomial of the order 8 (following  literature, see \citealt{pinna19,gadotti20,bittner20})\footnote{We note that we tried the fit for two different cases, one with only an additive Legendre polynomial and the other with only a multiplicative Legendre polynomial (both with order 8). We find no qualitative difference in the kinematic estimates and only a moderate variation in absolute measured velocity values of about 6\,km\,s$^{-1}$ (median value).}. Before the fitting, care was taken to mask wavelength ranges covered by the broad and bright emission-lines. We adopted a conservative approach and used broad masks with a width of up to 200 {\AA} for the broadest emission-lines (e.g., Ne~{\sc iv}, [Ar~{\sc iv}], H$\beta$, [O~{\sc iii}]$_{4959}$, [O~{\sc iii}]$_{5007}$, etc.; see Figure~\ref{fig: fits}) in order to ensure that these lines were masked even for the central spaxels with the broadest emission-lines. This ensured that no contamination from the emission-lines affected the analysis of the stellar kinematics, but still left sufficient stellar absorption features to obtain stable solutions for the stellar kinematics. We note that on reducing the masking width to 100\,{\AA} for the strongest emission lines, the velocity measurements are consistent in the outer Voronoi bins where the strongest emission lines are much narrower, giving us confidence the fitting wavelength coverage is sufficient.  In addition, the wavelength range from 5820-5970 \AA \, is masked, to avoid the contamination of the spectra from the laser guide stars of the adaptive optics facility. The stellar fits were visually inspected for each bin and were seen to closely follow the observed spectra in each case. As an example illustration, the observed spectra and the fits to the stellar continuum are shown in the lower panel of Figure \ref{fig: fits} (see supplementary material for the fits from all other Voronoi bins). In Section \ref{sec: results}, we discuss the resulting stellar kinematics. The median of errors, across all Voronoi bins, on the stellar kinematics were $\pm$\,16\,km\,s$^{-1}$ for stellar velocities and $\pm$\,18\,km\, s$^{-1}$ for stellar dispersion estimates, where the errors are formal errors (1$\sigma$) uncertainties estimated from \texttt{pPXF}.

We used the redshift estimate from the stellar fits of each Voronoi bin and estimated a flux-weighted mean redshift of $z\,=\,0.15004$. This value has then been used as the systemic redshift, for all of the analysis in this paper.

\subsection{Emission-line profile fitting procedure}\label{sec: emline}

In this work, we characterise the spatially-resolved kinematics and flux ratios of strong optical emission-lines (using the MUSE data) and the CO(3--2) emission-line (using the ALMA data). We take the same overall approach for fitting all emission-lines, which we describe first, before giving details that are specific to fitting individual emission-lines. All of the line profiles and velocity maps displayed in this work have been shifted to the rest frame, using the stellar systemic redshift described in Section \ref{sec: gist}. A set of emission-line profiles and fits from across the whole galaxy are shown in the supplementary material.

We fitted the profiles using the $\chi^2$ minimisation procedure {\sc curve\_fit} (\citealt{scipypaper}) and the most appropriate number of Gaussian components (detailed below) were selected using the Akaike Information Criteria (AIC; \citealt{akaike74}). Due to the complexities of the optical emission-line profiles, we primarily use these fits to obtain an overall non-parametric characterisation of the line profiles (see Section~\ref{sec: nonparam}) and largely refrain from assigning physical significance to individual components \citep[following e.g.,][]{liu2013,harrison14}. However, we verified that the non parametric results on the gas kinematics provide qualitatively the same conclusions as investigating `broad' and `narrow' components separately.

For our primary tracer of the ionised gas kinematics we use the [O~{\sc iii}]$\lambda$5007 emission-line. We fitted the spectral region around the [O~{\sc iii}]$\lambda\lambda$4959,5007 emission-line doublet using the rest-frame air wavelengths in the range [4940, 5040]\,\AA. We attempted to fit the [O~{\sc iii}]$\lambda$5007 profile using one, two or three Gaussian components (selected based on the AIC). An example [O~{\sc iii}]$\lambda$5007 (hereafter [O~{\sc iii}]) emission-line profile and the best-fit solution is shown in Figure~\ref{fig: fits}. We used the same Gaussian component parameters to fit simultaneously the [O~{\sc iii}]$\lambda$4959 emission profile, by using the same line width, a fixed wavelength separation and a fixed intensity ratio of [O~{\sc iii}]$\lambda$4959/[O~{\sc iii}]$\lambda$5007 = 0.33 \citep[e.g.,][]{dimitrijevic2007}.

For the purpose of obtaining estimates of electron densities (see Appendix \ref{app: ne}), we followed a similar procedure to fit other optical emission-lines (in all cases using rest-frame air wavelengths). We note that we used a simple first-order polynomial to characterise the local continuum around all the optical emission-lines. This is typically appropriate for the strong emission-line dominated spectra that we have and hence will not affect the inferred ionised kinematics. 

To trace the molecular gas kinematics, the CO(3--2) line (with a rest frame frequency of 345.7960\,GHz) was extracted from the ALMA data cube around a spectral window of $\sim$\,$\pm$\,700 km\,s$^{-1}$. Based on visual inspection and an AIC estimate, a maximum of two Gaussian components were required to characterise the CO(3--2) emission-line profile over the whole galaxy. The local baseline continuum was well characterised with a constant value. An example CO(3--2) emission-line profile and the best fit solution is shown in Figure~\ref{fig: fits} and a set of emission-line profiles and fits from across the whole galaxy are shown in the supplementary material.

\subsection{\textbf{Non-parametric velocity definitions}}\label{sec: nonparam}

The complexities and degeneracies of the emission-line fits exhibited in powerful quasars lead us to favour non-parametric definitions of bulk velocity shifts and velocity-dispersion \citep[also see][for a kinematic study of other QFeedS targets]{harrison14,jarvis2019}. Therefore, we used the best-fit models of the emission-line profiles from Section \ref{sec: emline} to calculate the following: 

\begin{itemize}
\item \textbf{$V_{50}$}, which is the median velocity of the overall emission-line profile, and is a tracer of the bulk gas velocity.
\item \textbf{$W_{80}$}, which is the velocity-width of the emission-line that contains 80 per cent of the integrated flux and is a tracer of the velocity-dispersion. This is defined as,
   $W_{80} = V_{90}-V_{10}$,  where $V_{10}$ and $V_{90}$ are the 10th and 90th percentiles of the emission-line profile, respectively. For a single Gaussian profile, $W_{80}$ is equivalent to 1.083$\times$FWHM or 2.6$\times \sigma$. 
\item Peak SNR ratio, which is simply the ratio of the peak flux density in the emission-line profile to the noise in the spectra. The noise is calculated by taking  the standard deviation of the continuum spectra, in windows of $\sim$ 30 \AA,\,on both sides of the emission peak. 
\end{itemize}

We note that we corrected the $W_{80}$ values by first subtracting, in quadrature, the line-spread function (LSF) $\sigma_{\mathrm{LSF}}$, from each of the component gaussians. The wavelength-dependent $\sigma_{\mathrm{LSF}}$ of the MUSE data was estimated using equation 8 from \texttt{udf-10} parametrisation of \cite{bacon2017}. 

In addition to tracing these non-parametric values, we also attempted to trace the properties of the individual broad and narrow components. Although these results are more subject to degeneracies, and therefore provided noisier results, we found that we draw the same conclusions on the spatial distribution of the high velocity-dispersion gas and positions of bulk velocity offsets discussed in Section~\ref{sec: res_gal} and Section~\ref{sec: res_nuc}. This can be visually verified by looking at the individual Gaussian components of the fits shown in the supplementary material.

To map the ionised and molecular gas kinematics and the emission-line flux ratios on the galaxy-scale, we use the same spatial Voronoi bins as obtained for the stellar kinematics maps described in Section~\ref{sec: gist}. This approach is sufficient for our goal of establishing a broad description of the galaxy kinematic structure and performing a like-for-like comparison between the stellar kinematics and the two gas phases (see Section~\ref{sec: res_gal}). Furthermore, we find that the gas is predominantly located within the extent of the stellar emission (see Figure~\ref{fig: J1316}; Section~\ref{sec: galaxy_overview}). We note that for the CO(3--2) gas, some bins had low SNR (i.e., $<$\,3; see Figure \ref{app: voronoi}), and are hence not shown for the ALMA maps.  A study of any additional faint gas in the circumgalactic medium is deferred to future work. To plot the values extracted from the fits in the maps, on the galaxy-scale, we required a signal-to-noise ratio (SNR$\geq$3). However, this only affected the galaxy-scale maps of CO(3--2) where the SNR was found to be lower than this threshold in three bins.  Maps of $V_{50}$ and $W_{80}$ for [O~{\sc iii}] and CO(3--2) are discussed in Section \ref{sec: res_gal} and \ref{sec: res_nuc}. A visual inspection of the data reveals that our maps provide a reliable and meaningful representation of the underlying data.

We also produced maps of the emission-line properties on the central-scale around the jets (see yellow box in Figure~\ref{fig: J1316}). For the central-scale maps, we performed the emission-line profile fits, following Section~\ref{sec: emline}, in spatial regions of 0.2$\times$0.2\,arcsec. Again, we required a SNR$\geq3$ to plot the derived quantities in these maps. Central-scale maps of SNR, $V_{50}$ and $W_{80}$ for both [O~{\sc iii}] and CO(3--2) are presented and discussed in Section \ref{sec: res_nuc}. We note that we lack sufficient SNR to map the stellar kinematics using these smaller spatial bins on this scale. For all the fitted parameters, we obtained the one standard deviation errors as \( \sqrt{diag(pcov)} \), where pcov is the estimated covariance of each parameter. To estimate the uncertainty in $W_{80}$, we used the error in the FWHM of the broadest gaussian component and for the uncertainty in $V_{50}$ we used the error in the velocity estimate of the narrowest gaussian component.

\begin{landscape}
 
 \begin{figure}
     \begin{center}
	\includegraphics[width=1.3\textwidth,height=0.7\textheight]{ 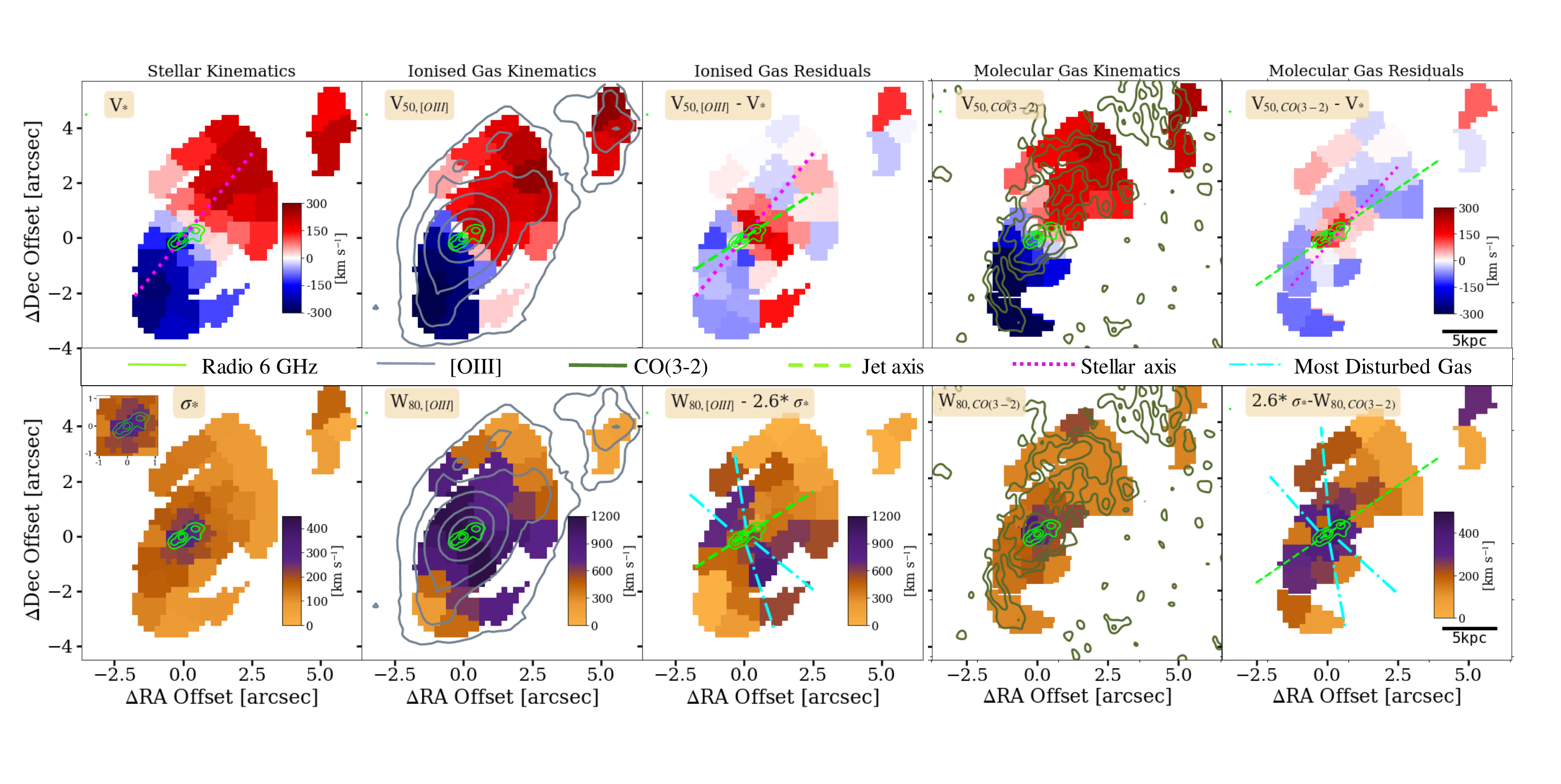}
    \caption{A comparison of the stellar and multi-phase galaxy-scale kinematics using the Voronoi bins. {\em Top row:} bulk velocity maps ($V_{\star}$ for the stellar and $V_{50}$ for the gas). {\em Bottom row:} velocity-dispersion maps ($\sigma_{\star}$ for the stellar kinematics and $W_{80}$ for the gas). In both rows, {\em from left-to-right:} stellar kinematics, ionised gas kinematics (from [O~{\sc iii}]), residual ionised gas kinematics (i.e., the velocity difference between the stellar and ionised gas maps; see Section \ref{sec: res_gal_ion}), molecular gas kinematics (from CO(3--2)), and residual molecular gas kinematics. In all panels, the light-green contours represent the 6\,GHz radio emission at levels [32,16,4]\,RMS$_{\mathrm{radio}}$.  Grey contours (in second column) trace the [O~{\sc iii}] emission at [64,16,4,2,1]\,RMS$_{\mathrm{[OIII]}}$. The dark-green contours trace the CO(3-2) emission at levels of [16,8,4,2,1]\,RMS$_{\mathrm{CO(3-2)}}$. The dashed green and dotted pink lines trace the projected jets and stellar kinematic axis, respectively. A scale bar representing a length of 5\,kpc is shown in both the right-most panels. The bulk of the gas follows the stellar velocities at the galaxy scale as seen in both the phases, except for the central bins, around the jet region, which show strong velocity residuals (top-row: third and fifth column). The highest dispersion values are seen perpendicular to the jets for ionised gas (bottom-row, third column), indicated by the cyan-lines (traced from the cones of highest velocity-dispersion identified in Figure \ref{fig: muse_zoom}).}
    \label{fig: muse_alma_gist}
     \end{center}
\end{figure}
\end{landscape}

\begin{figure*}
\centering
\includegraphics[width=1\textwidth]{ 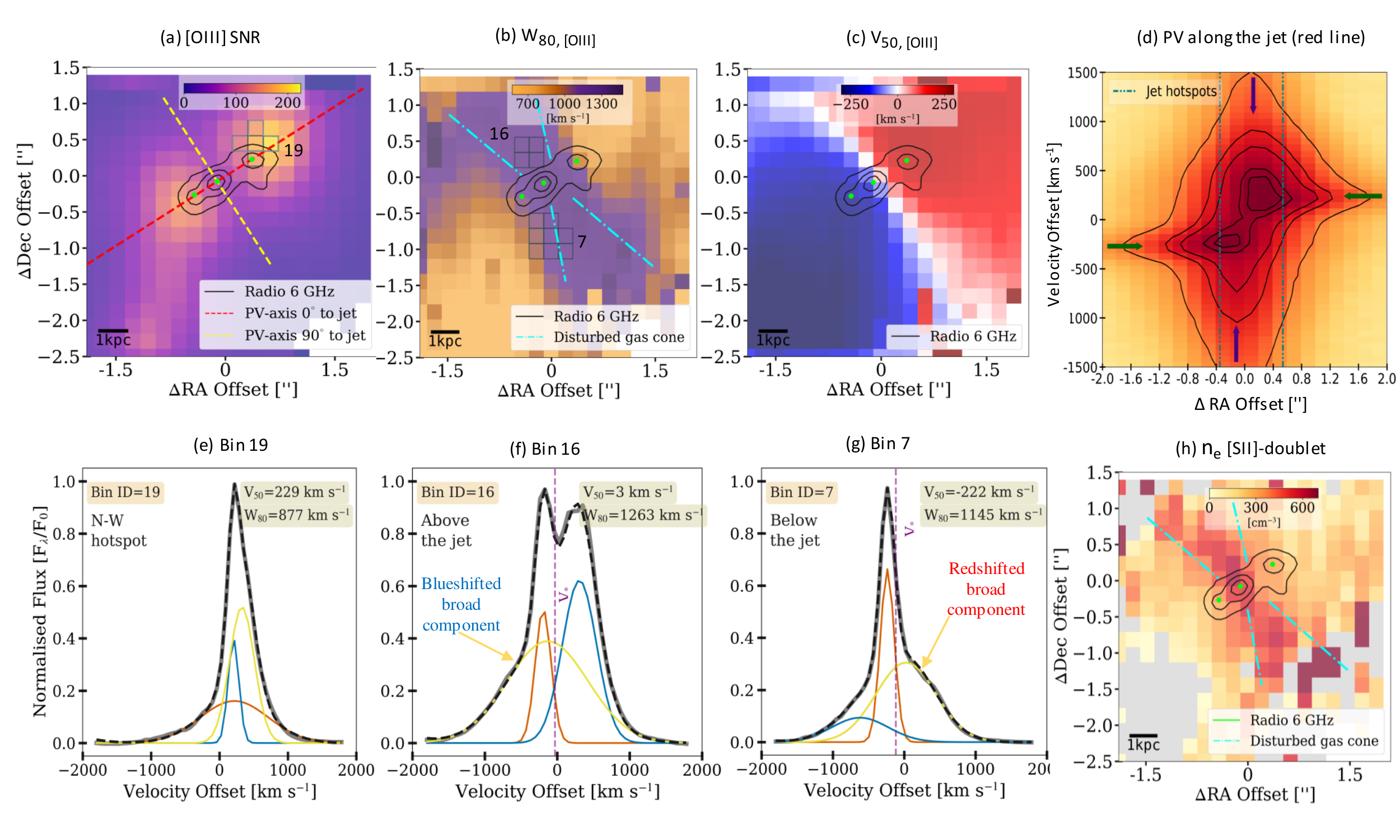}
\caption{Kinematic analysis of the ionised gas in the region around the central jet, using maps of the non-parametric values of the [O~{\sc iii}] emission-line fits and position velocity diagrams around the [O~{\sc iii}] emission-line (see Section \ref{sec: res_nuc_ion}). The black contours in each map represent the 6 GHz radio emission at levels of [32,16,4]\,RMS$_{\mathrm{radio}}$. The three cyan dots represent the HR:A (core) and jet hot spots (HR:B and HR:C), as in Figure \ref{fig: J1316}. A 1\,kpc scale bar is shown in each of the maps. \textbf{Panel a:} Peak-SNR map; \textbf{Panel b:} $W_{80}$ map; \textbf{Panel c:} $V_{50}$ map; \textbf{Panel d:} PV diagram extracted along the jet axis (as shown through the dashed red axes in panel {\em Panel a:}), with the position-offset along x-axis and velocity-offset along y-axis, and the vertical lines indicate the locations of the radio jet hot spots; \textbf{Panel e:} the [O~{\sc iii}] line profile from regions of highest peak-SNR values; \textbf{Panel f and Panel g:} the [O~{\sc iii}] line profiles from regions of disturbed gas, above (Voronoi bin 16) and below (Voronoi bin 7) the jet, respectively,  and \textbf{Panel h:} electron density map from [S~{\sc ii}] doublet. The gaussian components (in \textit{panels, e, f, and g}) are as described in Figure \ref{fig: fits}. In {\em Panel b}, the emission corresponding to an enhanced velocity-dispersion ($W_{80}$ $\geq$1000 km s$^{-1}$), suggests a bi-conical structure in the region perpendicular to the jets (indicated by the dash-dotted cyan lines). The PV diagram in {\em Panel d}, reveals enhanced velocities behind the jet hot spots, as highlighted by the purple vertical arrows. Narrow velocity components, highlighted with horizontal darkgreen arrows in {\em Panel d}, are observed to propagate roughly from the radio hotspots and extend for 0.8\,arcsec (2.1\,kpc) in each direction.}
\label{fig: muse_zoom}
\end{figure*}


\section{Results}\label{sec: results}
In this section we present our results from the analysis of MUSE and ALMA data for the $z$\,=\,0.15, type-2 quasar, J1356+1753, selected from QFeedS (see Figure~\ref{fig: mullaneyplot}). Specifically, we present: (1) the distribution of stellar emission, ionised gas and molecular gas (Section~\ref{sec: galaxy_overview}); (2) the galaxy-scale stellar and multi-phase gas kinematics (Section~\ref{sec: res_gal}) and (3) the central-scale gas kinematics and emission-line properties close to the jets (Section~\ref{sec: res_nuc}). 

\subsection{Overview of the stellar and gas distribution}\label{sec: galaxy_overview}

Figure \ref{fig: muse_alma_gist} presents the galaxy-scale stellar kinematics (first column), ionised gas kinematics (second column), residuals by subtracting the stellar and ionised gas kinematics  (third column), the molecular gas kinematics (fourth column) and the residuals from subtracting the molecular and stellar kinematics (fifth column). In all columns, the upper-row presents the bulk velocity maps (V$_{\star}$ or $V_{50}$) and the lower-row presents the velocity-dispersion maps ($\sigma_{\star}$ or $W_{80}$) maps. We note that while computing residuals, the stellar dispersion values were scaled by a factor of 2.6, to make the values more directly comparable to $W_{80}$, which we used to characterise the velocity width of the gas (see Section~\ref{sec: nonparam} for the estimation methods). 

As can be seen in Figure~\ref{fig: J1316}, J1316+1753 is a spiral galaxy extending over $\sim$\,20\,kpc ($R_{1/2}$\,=\,6.1\,kpc) from the north spiral arm to the south spiral arm. [O~{\sc iii}] emission is our tracer of the ionised gas. The [O~{\sc iii}] image shown in the left-most panel of the figure reveals that the ionised gas broadly covers the same spatial regions as the stellar continuum emission in both the primary galaxy and the minor companion galaxy to the north west (also see the second column of Figure~\ref{fig: muse_alma_gist}). However, we also observe an ionised gas cloud (i.e., with no corresponding stellar continuum emission), at a distance of $\sim$\,17\,kpc north west of the galaxy nucleus, with a blue-shifted velocity of -\,250 km\,s$^{-1}$. Although not the focus of this work, we speculate that this cloud may be a tidal debris or the result of an historic outflow (see \citealt{lintott09,keel12,keel19} and the references therein). 

CO(3--2) emission is our tracer of the molecular gas (see Section \ref{sec: datared_alma}). The CO(3--2) contours shown in the right-most panel of Figure~\ref{fig: J1316} and in the fourth column of Figure~\ref{fig: muse_alma_gist} reveal that the molecular gas broadly follows the distribution of the stellar continuum emission, including tracing the northern and southern spiral arms. However, at the sensitivity of our observations, the molecular gas is less extended than both the [O~{\sc iii}] and continuum emission. Specifically, CO(3--2) emission is not detected towards the outer edges of the spiral arms. Nonetheless, for this work we only aim to qualitatively describe the galaxy-scale molecular gas kinematics, for which the data is sufficient.

\begin{figure*}
\centering
\includegraphics[width=1\textwidth]{ 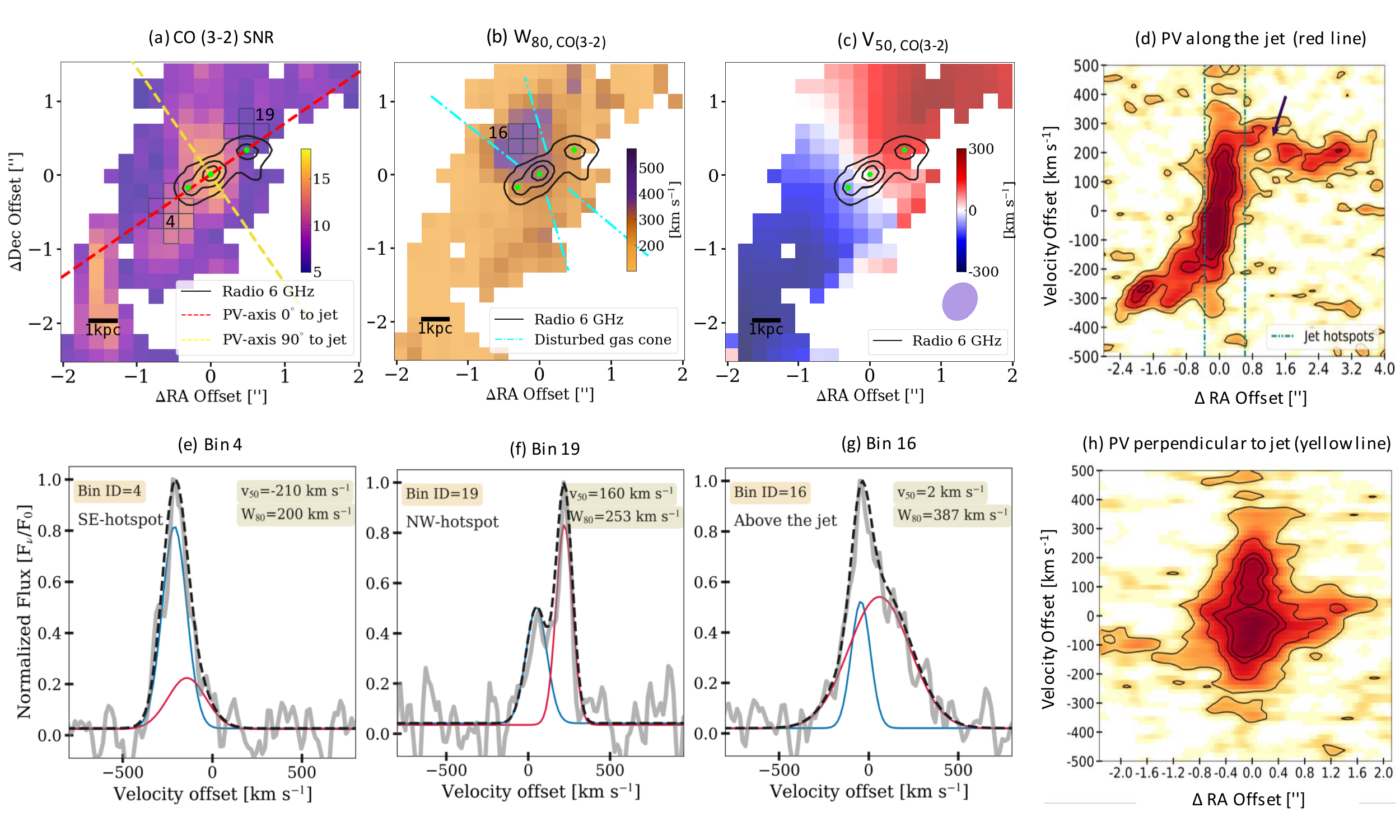}
\caption{Kinematic analysis of the molecular gas in the region around the jet, using maps of the non-parametric values of the CO(3--2) emission-line fits and position velocity (PV) diagrams around the CO(3--2) emission-line (see Section \ref{sec: res_nuc_mol}). The black contours in each of the maps represent the 6 GHz radio emission at levels of [32,16,4]\,RMS$_{\mathrm{radio}}$. \textbf{Panel a:} Peak-SNR map; \textbf{Panel b:} $W_{80}$ map; \textbf{Panel c:} $V_{50}$ map; \textbf{Panel d:} PV diagram extracted along the jet axis (see red dashed axes in {\em Panel a:}), with the position-offset along x-axis and velocity-offset along y-axis, and the vertical lines indicate the hot spots of the radio jets; \textbf{Panel e:} and \textbf{Panel f:} CO(3--2) line-profiles from the locations of the jet hotspots (Voronoi bin 4 and 19); \textbf{Panel g:} CO(3--2) line-profile from region of the broad disturbed gas just above the jet (Voronoi bin 16) and \textbf{Panel h:} PV diagram extracted along an axis perpendicular to the jets (see yellow dashed axis in {\em Panel a:}). The components in the line-fit plots are as described in Figure \ref{fig: fits}. {\em Panel b:} shows the presence of disturbed molecular gas in a localised region perpendicular to the jet (Voronoi bin 16). The PV diagram in Panel d, reveals a break in the velocity profile at the north-west jet hotspot (HR:C, highlighted with an arrow).}
\label{fig: alma_zoom}
\end{figure*}

\subsection{Kinematics on the galaxy-scale}\label{sec: res_gal}
As a first investigation of the stellar and multi-phase gas kinematics, we look at the entire galaxy-scale (i.e., across a 26$\times$26\,kpc region; see Figure~\ref{fig: J1316}) using the Voronoi bins defined in Section~\ref{sec: gist} (also see Figure~\ref{fig: voronoi_map}). By using the same spatial bins, we are able to directly compare the stellar, [O~{\sc iii}] and CO(3--2) kinematics without having to rely on a particular dynamical model with the associated uncertain assumptions. We note that for the following results, we are only concerned with the contiguous Voronoi bins of the central galaxy and we exclude the minor companion galaxy when presenting quantitative values. Our goal here is to qualitatively assess if the multi-phase gas follows the host galaxy's gravitational motions, before performing a deeper investigation of the kinematics and distribution of the gas in the central region of the galaxy in Section~\ref{sec: res_nuc}.

\subsubsection{Stellar kinematics}\label{sec: stelkin}
The stellar velocity and stellar velocity-dispersion maps, presented in Figure~\ref{fig: muse_alma_gist}, clearly suggest the presence of a smoothly rotating stellar galaxy disk. We obtain an approximate major kinematic axis of 140\,$\degree$\, by connecting the Voronoi bins with the highest values of projected velocities (i.e., $-$\,268\,$\pm$\,18\,km\,s$^{-1}$ and $+$\,240\,$\pm$\,14\,km s$^{-1}$) through the galaxy nucleus. This major kinematic axis is well aligned with the morphological axis of 150\,\degree, which we measured in Section~\ref{sec: angle_size}. Accounting for the 46\degree\, inclination angle of the galaxy (see Section~\ref{sec: angle_size}), we estimated the de-projected maximum rotation velocities of $-$\,360\,km\,s$^{-1}$ and $+$\,330\,km s$^{-1}$. These high rotation velocities are consistent with previous work suggesting that this galaxy has a very high mass \citep[i.e., the inferred stellar mass from SED fitting is $=$ 10$^{11\pm0.3}$\,M$_{\odot}$;][]{jarvis2020}. 

The observed stellar velocity-dispersion ranges from 81\,$\pm$\,22\,km\,s$^{-1}$ in the outer spiral arms to much higher values between 290\,--\,360 km s$^{-1}$ in the very central regions, with a median value of 326\,$\pm$\,27\,km\,s$^{-1}$ (see small inset in Figure \ref{fig: muse_alma_gist}). We discuss the interesting origin of the highest central stellar velocity-dispersion, and the remarkable spatial alignment with the radio jets in Section~\ref{sec: discussions}.

\subsubsection{Ionised gas kinematics}\label{sec: res_gal_ion}

In Figure~\ref{fig: muse_alma_gist} (second column) the kinematic structure of the bulk ionised gas velocity on large scales, i.e. $V_{\rm 50, [O III]}$, can be seen to broadly follow the stellar velocity structure. Indeed, the residuals (third column) created by subtracting the stellar kinematic values from $V_\mathrm{{50,[O III]}}$, reveals only modest residuals in the majority of the bins outside of the central regions, with $|V_{\rm 50,[O III]}-V_{\star}|~\leq$\,50\,km\,s$^{-1}$ in the spiral arms. We therefore infer that on large scales the [O~{\sc iii}] emission is broadly following the gravitational motions of the host galaxy. In contrast, the residuals between V$_{\rm 50,[O III]}$ and V$_{\star}$ are much larger in the central, $\lesssim$10\,kpc regions of the galaxy, reaching values of up to $-$\,105\,km\,s$^{-1}$ and $+$\,260\,km\,s$^{-1}$. We note that the bin at the tip of both  spiral arms also shows a large velocity residual of $=$\,139\,km\,s$^{-1}$ that we attribute to disturbed ionised gas extending as far as these locations (see below). 

A comparison between the stellar velocity-dispersion and velocity-dispersion of the [O~{\sc iii}] emission further confirms non-gravitational ionised gas motions (Figure~\ref{fig: muse_alma_gist}). The $W_{\rm 80,[O III]}$ map reveals extreme ionised velocity widths over very large scales. With the exception of the outermost north west and south east regions, the majority of the bins have [O~{\sc iii}] velocity widths much greater than the local stellar velocity-dispersion. Most dramatically,  perpendicular to the jet axis, high velocity dispersion is seen with W$_{80}>>2.6\,\sigma_{\star}$ (with residuals of typically 740\,km\,s$^{-1}$). The velocity-dispersion in these bins ranges from 1000-1300\,km s$^{-1}$, and hence we identify this as the most disturbed gas with W$_{80}\geq$1000\,km\,s$^{-1}$. This disturbed gas is seen to extend as far as the outer Voronoi bins which are located 7.5\,kpc each side of the nucleus. While the disturbed gas may be extending beyond this distance, we take a conservative estimate of 7.5\,kpc, obtained from the Voronoi bins defined at a SNR\,>\,5 (as explained in Section \ref{sec: gist}). Therefore we conclude that, whilst the [O~{\sc iii}] gas follows the large-scale rotation of the host galaxy, we see significant non-gravitational motions around the jets that is detected to a projected distance of 7.5\,kpc along the minor axis of the host galaxy.

 For an estimate of the outflow velocities, we obtained the maximum (terminal) velocities following \cite{heckman04}, as \(V_{\mathrm{max}}\,=\,\Delta\,V\,-\,FWHM_{\mathrm{broad}}/2\), where \(\Delta\,V\,=\,V_{\mathrm{broad}}-V_{\star} \) (also see e.g., \citealt{rupke05,veilleux05}). We found these velocity estimates\footnote{V$_{\mathrm{max, Bin 28}}$\,=\,480\,$\pm$\,17\,km\,s$^{-1}$; V$_{\mathrm{max, Bin 33}}$\,=\,650\,$\pm$\,10\,km\,s$^{-1}$} in the terminal bins (precisely, Voronoi bin 28 and 33) to be similar to our non-parametric estimate of $W_{80}$\footnote{W$_{\mathrm{80,Bin 28}}$\,=\,880\,$\pm$19\,km\,s$^{-1}$; W$_{\mathrm{80,Bin33}}$\,=\,820\,$\pm$26\,km\,s$^{-1}$} in the same location and hence we decided to use the $W_{80}$ values for our estimate of the outflow velocity in Section \ref{sec: energetics} (also see e.g., \citealt{liu2013,harrison14}). This analysis of the galaxy scale ionised gas kinematics reveals non-gravitational motions in and around the nuclear radio jets in J1316+1753, with high velocity-dispersion gas extending to a distance of 7.5\,kpc each side of the nucleus. We investigate the velocity structure of the central regions in more detail in Section~\ref{sec: res_nuc}. We note that the spatially-resolved emission-line ratio diagnostic for the dominant ionising source reveals AGN to be the dominant ionising source across all galaxy bins (except in the companion galaxy), however, we defer the detailed discussions to future works.

\subsubsection{Molecular gas kinematics}\label{sec: res_gal_mol}

Using the kinematics of CO(3-2) as a tracer in Figure~\ref{fig: muse_alma_gist}, we find that the molecular gas shows a similar, but less extreme, behaviour than the ionised gas. The bulk velocity of the molecular gas on large scales broadly follows that of the stellar velocity. The peak projected velocities of the CO(3--2) are similar to the peak stellar velocities and furthermore, the major kinematic axes are found to be around 140\,\degree\, for both the molecular and stellar kinematics. The $|\mathrm{V}_{\rm 50,CO(3-2)}-\mathrm{V}_{\star}|$ residuals are small (i.e., $\leq$70\,km\,s$^{-1}$) in all but the central 6 Voronoi bins\footnote{Specifically, the Voronoi bins 0, 1, 2, 13, 17, 19 (see Figure \ref{fig: voronoi_map})}, which on average have residuals of 190\,km\,s$^{-1}$. Therefore, the bulk velocity of the molecular gas is even closer to the stellar kinematics than found for the ionised gas, but still shows significant non-gravitational motion in regions around the radio jets. This is investigated further in Section~\ref{sec: res_nuc}.  

Figure \ref{fig: muse_alma_gist} shows that, similar to the ionised gas, the molecular gas has a velocity-dispersion comparable to the stellar kinematics  in the outer bins, but with much higher residuals (i.e., 110 km s$^{-1}$) in the central bins. However, whilst the central 7 Voronoi bins\footnote{Specifically, the Voronoi bins 0, 1, 2, 3, 13, 16, 17 (see Figure \ref{fig: voronoi_map})} show the highest velocity widths of W$_{\mathrm{80,CO(3-2)}}$ =\,380-500\,km\,s$^{-1}$, these are on average, a factor of 3 less than those seen in [O~{\sc iii}] at the same location. The most distant spaxel where this disturbed molecular gas is observed (i.e., with W$_{80}>$380\,km\,s$^{-1}$) in Figure \ref{fig: muse_alma_gist} has a projected distance from the radio core of 2.5\,kpc. Therefore, the spatial extent of the high velocity-dispersion molecular gas is seen to be approximately a third of that seen in the ionised gas, i.e. 2.5\,kpc\, (Figure~\ref{fig: muse_alma_gist}). We investigate this further by zooming into the central regions in the following subsection.

\subsection{Kinematics on the central-scale}\label{sec: res_nuc}

Following the identification of non-gravitational gas motions identified in the central region of J1316+1753, presented in the previous sub-section, we were motivated to map the ionised and molecular gas properties in more detail in this region. To do this, we mapped ionised and molecular gas kinematics across the central, 10$\times$10\,kpc region (see Section~\ref{fig: J1316}) using smaller spatial bins, and with a SNR cut of $\geq$\,3. For the central-scale maps, we performed the emission-line profile fits, in spatial regions of 0.2$\times$0.2\,arcsec, i.e., in individual spaxels for the MUSE data and in bins of 4$\times$4 pixels for the ALMA data to have a comparable resolution as MUSE. The results are described below and are illustrated in Figure \ref{fig: muse_zoom} for ionised gas and in Figure~\ref{fig: alma_zoom} for the molecular gas. As discussed in Section~\ref{sec: gist} we lack the sensitivity to fit the stellar continuum with finer binning; however, in Section~\ref{sec: positivefeedback} we discuss the remarkable alignment between the jets and stellar emission and kinematics in this region using the larger spatial bins.

\subsubsection{Ionised gas around the jet}\label{sec: res_nuc_ion}
 
In Figure~\ref{fig: muse_zoom}, we show maps of SNR, $W_{80}$, $V_{50}$ and $n_e$, along with some line-profiles for the [O~{\sc iii}] emission (see Figure caption for detail). Contours of radio emission are overlaid, showing the relative location of the radio jets. Particularly striking is the high levels of velocity-dispersion running perpendicular to the jet axis, apparent in the $W_{80}$ maps shown in panel (b). The high velocity-dispersion in the ionised gas is further supported by the position-velocity (PV) diagram shown in panel (d), which was constructed by summing over a 0.6\,arcsec-wide pseudo-slit, along the jet axis. It is quite clear that these central regions show a broad velocity component with maximum velocities of around $\pm$1500\,km\,s$^{-1}$.

Indeed, a visual inspection of the [O~{\sc iii}] emission-line profile just north and just south of the jets (see panel (f) and (g), respectively), reveals a very broad component in these regions, with FWHM\,$=$\,980\,$\pm$\,46\,km\,s$^{-1}$ and 1310\,$\pm$\,18\,km\,s$^{-1}$, respectively. North of the jet, this component is blueshifted by 100\,km\,s$^{-1}$ with respect to the local stellar velocity and south of the jet this component is redshifted by 140\,km\,s$^{-1}$. We note that this shows the opposite red- and blue-shifts with respect to the galaxy rotation. This highest velocity-dispersion emission (with W$_{80}\,\geq$\,1000\,km\,s$^{-1}$) starts from just behind the jet hot spots (highlighted with green dots in the $W_{80}$ map and vertical dot-dot-dashed lines in the PV-diagram). 

All of these observations suggest outflowing, high velocity-dispersion gas that originates from behind the jet hot spots and expands, in what appears to be a bi-cone structure. We trace the outer-edges of these cones by drawing straight dot-dashed lines, on the $W_{80}$ map for spaxels along the axis perpendicular to the jet and identified with $W_{80}\geq$1000\,km\,s$^{-1}$ to identify the morphology of the most extreme gas kinematics (see Section \ref{sec: res_gal_ion}). These cones have opening angles of $\sim$\,40\,\degree. To understand how this translates to the observations we made on the galaxy scales (Section~\ref{sec: res_gal}), we overlaid them onto Figure~\ref{fig: muse_alma_gist}. Re-reassuringly, this gives a consistent picture because the highest velocity-dispersion gas seen on the galaxy-scale in Figure~\ref{fig: muse_alma_gist}  (i.e., out to 7.5\,kpc each side of the nucleus), is located within these cones.

In panel (h) we show the electron density in the individual spaxels as estimated by using the ratio of the [S~{\sc ii}] doublet (see Appendix \ref{app: ne} for more details and caveats). This reveals that the spaxels with the highest inferred electron density (i.e., 500-600 cm$^{-3}$) are co-spatial with the cones of highest velocity-dispersion ionised gas. We verified that this behaviour continues onto the galaxy scale by creating a map with electron density for each of the Voronoi bins and present the results in Appendix \ref{app: ne}. The spaxels outside the cones of highest velocity-dispersion regions have more modest electron densities ($\leq$ 150 cm$^{-3}$). We discuss the implications of these observations in Section \ref{sec: jet_ism}. 
 
Another key observation of the [O~{\sc iii}] emission in the central regions, presented in Figure~\ref{fig: muse_zoom}, is that the highest SNR emission is concentrated in two regions; one around each of the two radio jet hot spots. A visual inspection of the emission-line profiles demonstrates that this is due to two narrow emission-line components that are particularly strong in each of these locations. These narrow emission-line components, which can be seen in the emission-line profiles shown in panel (e) and panel (g) of Figure~\ref{fig: muse_zoom}, are separated in velocity by 441\,$\pm$\,14\,km\,s$^{-1}$.
Indeed, the $V_{50}$ map, shown in panel (c) of Figure~\ref{fig: muse_zoom}, highlights this further; we observe a strong velocity gradient {\em directly} along the jet axis. We note that this is offset by 20\,\degree\, from the galaxy kinematic axis and we have already shown that the gas in these regions is not dominated by gravitational motions (Section~\ref{sec: res_gal_ion}). The PV diagram in panel (d) of Figure~\ref{fig: muse_zoom} shows that these two strong narrow emission-line components, which are highlighted with  horizontal arrows on the panel, are emanating with  velocities of $+$206\,$\pm$\,15\,km\,s$^{-1}$ and $-$235\,$\pm$\,1.3\,km\,s$^{-1}$ from the location of the two jet hot spots. These jet hot spot locations are highlighted with vertical lines in the panel. We discuss these ionised gas velocity components in the context of jet-ISM interactions in Section~\ref{sec: jet_ism}.

\subsubsection{Molecular gas around the jet}\label{sec: res_nuc_mol}

In Figure~\ref{fig: alma_zoom}, we present a similar set of figures for the CO(3--2) emission in the vicinity of the radio jets, as we did for the [O~{\sc iii}] emission in Figure~\ref{fig: muse_zoom}, namely, maps of SNR, $W_{80}$, and $V_{50}$, along with some line-profiles for the CO(3--2) emission (see Figure caption for detail). We observe the presence of molecular gas with high velocity-dispersion perpendicular to the jet axis (Figure \ref{fig: alma_zoom}; panel (b)). A visual inspection of the emission-line profile of this region (see panel (g)), shows that there is a broad CO(3--2) emission-line component (FWHM of 404\,$\pm$10\,km\,s$^{-1}$) responsible for the high $W_{80}$ values seen in this location. Furthermore, Panel (h) shows a PV diagram extracted along a 0.6\,arcsec slit perpendicular to the jet axis, which further confirms a region of high velocity-dispersion along this axis. 

The disturbed molecular gas is located within the same cone identified in the ionised gas phase (see dot-dashed lines on panel (b) of Figure~\ref{fig: muse_zoom}), but, this time, only in the northern side of the jet. Furthermore, the disturbance appears to be less extreme. Specifically, the FWHM of the broad [O~{\sc iii}] emission-line component in the same region is  3$\times$ higher than that observed in CO(3-2) at the same location. Additionally, the disturbed molecular gas is seen to extend to 2.5 kpc, which is only a third the extension of the disturbed ionised gas. 

In the PV diagram of CO(3--2), created using a 0.6\,arcsec pseudo-slit running along the jet axis (panel (d) in Figure~\ref{fig: alma_zoom}), we observe a small velocity jump of around 100\,km\,s$^{-1}$ at the location of the northern radio hot spot (highlighted with an arrow in the Figure). Further evidence for irregular kinematics in this region comes from a visual inspection of the emission-line profile (see panel (f) in Figure~\ref{fig: alma_zoom}) which shows that the profile is split into two components separated by 169\,$\pm$9\,km\,s$^{-1}$. Finally, there is some evidence for a deficit of CO(3--2) emission close to the northern hot spot of the jet, as indicated by a black arrow in the figure. From the PV-diagram, we estimate the SNR of the CO flux to be 11-14 times higher before and after this cavity and about 90\,\% less flux is seen in the cavity as compared to the surroundings. These observations coupled with the strong narrow emission-line components seen in the ionised gas profile, serve as evidence of possible jet-ISM interactions happening in this region.


\begin{figure*}
\centering
\begin{tabular}{cc}
\subcaptionbox{}{\includegraphics[width=.5\textwidth,height=8cm]{ 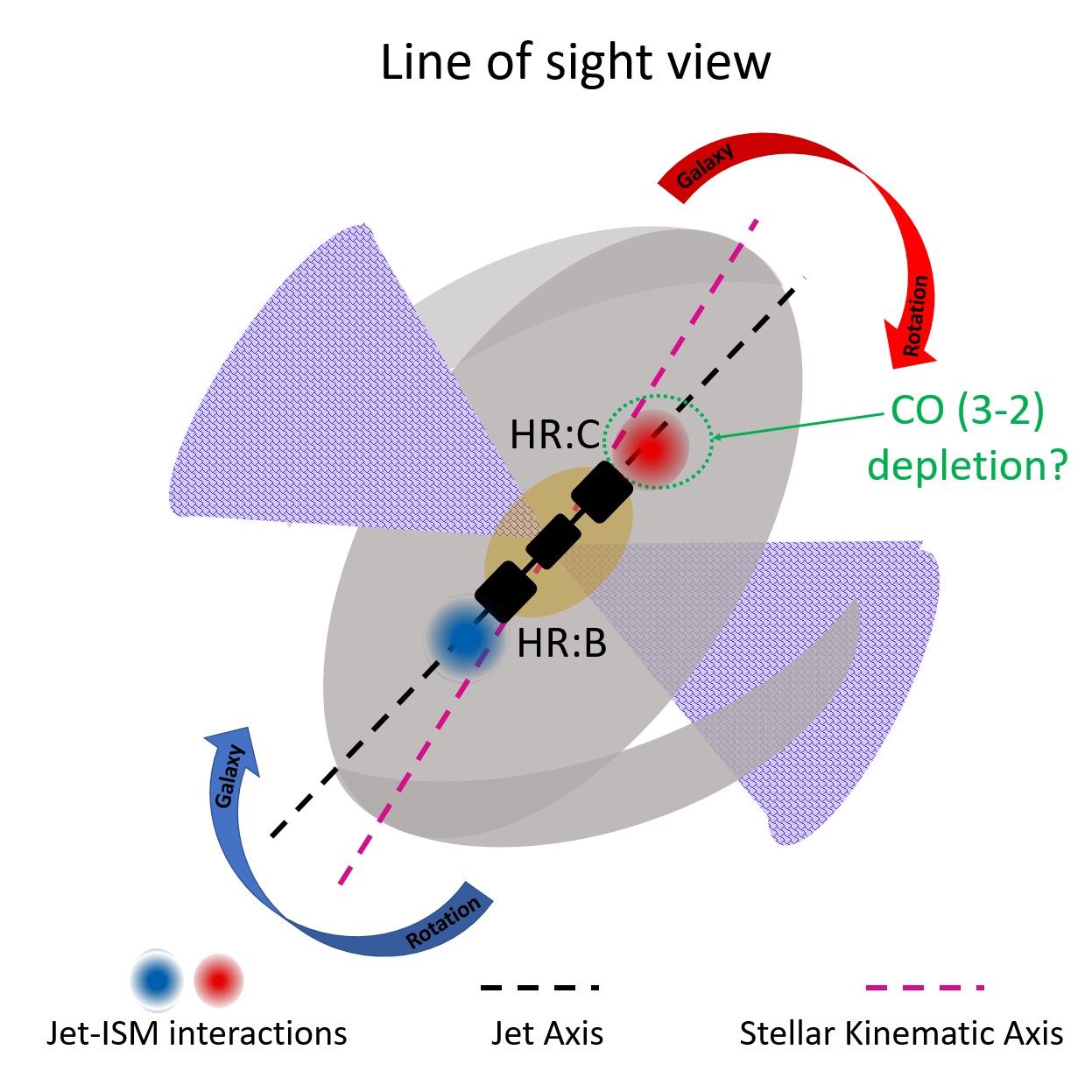}} & \subcaptionbox{}{\includegraphics[width=.5\textwidth,height=8cm]{ 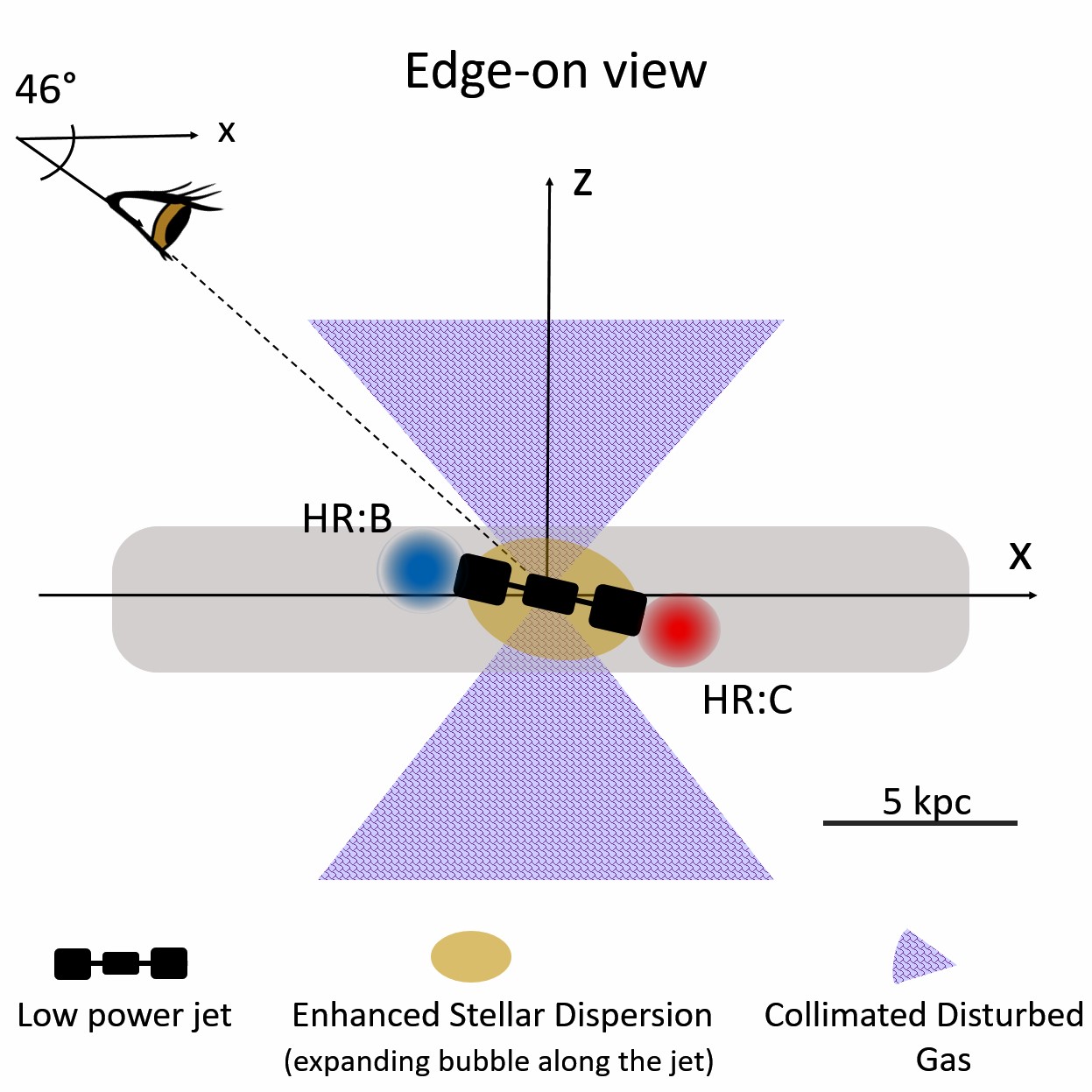}} \\ 
\end{tabular}
\caption{A schematic view of the galaxy J1316+1753 to highlight the key observations identified in this work. {\em Left:} The line-of-sight view (where North is up and East is to the left); {\em Right:} Our inferred edge-on view. A legend shows the different symbols used to represent the main features of interest and an approximate scale bar is shown in the right panel at the bottom. A detailed description of the features in this schematic is presented in Section \ref{sec: jet_ism}.}
    \label{fig: schematic}
\end{figure*}

\section{Discussion}\label{sec: discussions}

We have presented new MUSE and ALMA data for a type-2, $z\,=\,0.15$ quasar, J1316+1753, selected from the Quasar Feedback Survey (Figure \ref{fig: mullaneyplot}). Based on the previous radio data, this source contains low-power ($P_{\rm jet}$\,=\,10$^{44}$\,erg\,s$^{-1}$) $\sim$1\,kpc radio jets. These jets are confined within the galaxy disc and inclined at $\lesssim$\,5\,\degree\, with respect to the plane of the galaxy disc, with the south-west jet (HR:B) as the approaching side (see Section \ref{sec: jetangle} and Figure \ref{fig: J1316}). We summarise our key observations on the stellar, warm ionised gas and cold molecular gas kinematics with a schematic representation in Figure \ref{fig: schematic}. This is presented both from the line-of-sight point of view ({\em left-panel}) and an edge-on view ({\em right-panel}). Below, we discuss the observational features that appear to be tracing the interaction of the jets with the host galaxy's ISM (Section~\ref{sec: jet_ism}), before discussing the possible evidence for feedback effects on the host galaxy (Section~\ref{sec: feedback}) and then discussing the results in the context of the overall AGN population (Section \ref{sec: lit_comp}).

\subsection{Turbulent outflows and jet-ISM interactions}\label{sec: jet_ism}

Here we describe how our observations are consistent with the low power jets, which are highly inclined into the galaxy disk, are interacting with the ISM and driving large scale turbulent outflows.

We see evidence for jet-ISM interactions on the multi-phase gas at the jet termini (highlighted with blue/red regions in Figure \ref{fig: schematic}). Strong velocity enhancements in the warm ionised gas are seen, with two high velocity offset ionised gas components, separated by 441\,km\,s$^{-1}$, that are brightest in the regions of the jet termini and appear to propagate away along the jet axis (Figure \ref{fig: muse_zoom}). In the molecular gas phase, a $-$\,100\,km\,s$^{-1}$ change in the molecular gas velocity is observed just beyond the HR:C radio hot spot (Figure~\ref{fig: alma_zoom}). This is consistent with previous work that has found velocity jumps in multiple phases of gas at the locations of jet termini (e.g., \citealt{morganti13,riffel14,tadhunter14, morganti15,ramosalmeida17,finlez18, jarvis2019,husemann19,ruffa19,santoro20, morganti21}). Furthermore, our results confirm that the strongly double peaked [O~{\sc iii}] emission-line profile, seen in the SDSS spectrum of J1316+1753, is not the result of the presence of a dual AGN despite it being such a candidate (\citealt{xukomossa2009, smith2010, lyuliu2016, baronpoznaski2017}). Indeed, studies have also shown that double-peaked emission-lines could mostly be an indication of the existence of (jet-driven) galaxy-wide outflows (see \citealt{rosario10, kharb2015, nevin18,kharb2019,kharb21,RamosAlmeida22}).

We also observe tentative evidence for depleted CO\,(3--2) emitting gas at the location of the brightest jets (see Figure~\ref{fig: alma_zoom}). We indicate this with a green-dashed circle in the schematic in Figure \ref{fig: schematic} (also see top-right panel, Figure \ref{fig: alma_zoom}). This is the same region where we see a bright warm ionised gas component in the [O{~\sc iii}] (Figure~\ref{fig: muse_zoom}). Possible explanations for this include molecular gas removal or an interaction between the radio jets and the ISM that suppresses CO emission (e.g., through excitation or dissociation) while boosting the emission of the warm ionised gas (e.g., \citealt{Rosario2019,shimizu2019}). Obtaining more molecular gas tracers, such as multiple CO lines and rotational and vibrational H$_{2}$ lines would help to verify this observation and differentiate between these scenarios (e.g., \citealt{Rigopoulou02, Dasyra16}).

Further evidence for an impact of the jets on the host galaxy ISM comes from the very high velocity-dispersion ionised gas (W$_{80}\geq$1000\,km\,s$^{-1}$; i.e., $\sigma\sim$400\,km\,s$^{-1}$) originating from just behind the jet hot spots, and outflowing in an apparent bi-cone, in a perpendicular direction as far as project distance of 7.5\,kpc (see Figure \ref{fig: muse_alma_gist} and Figure ~\ref{fig: muse_zoom}). The regions of this high velocity-dispersion ionised gas are highlighted as the purple hatched regions in the schematic diagram in Figure \ref{fig: schematic}. This disturbed gas can be seen to be moving along the galaxy minor axis and due to the orientation of the galaxy is seen to be moving towards us (and hence blue-shifted) above the galaxy plane and moving away from us (and hence red-shifted) below the galaxy plane. We also observe that this turbulent gas has a three times higher electron density than the rest of the galaxy disk (see Figure \ref{fig: muse_zoom}). Such high electron densities within outflowing components have also been seen in previous studies \citep[e.g,][]{villarmartin14, perna17, mingozzi19, davies20, fluetsch21}. A possible explanation for this could be due to the compression of the gas expelled in the outflow \citep[e.g.,][]{bourne15,decataldo19}. 

In the molecular phase, we also see enhanced velocity-dispersion in the same direction as the turbulent ionised gas, but only in the regions north of the jet (not indicated in Figure \ref{fig: schematic} for simplicity) and extends to about a distance of 2.5\,kpc. A possible explanation for the less extreme kinematics observed in CO, compared to [O~{\sc iii}], could be that the low velocity gas has cooled more efficiently, as the higher velocity gas has a higher post-shock temperature (e.g. \citealt{costa15}). The densest material in the ISM will also naturally be driven to lower velocities, which would also result in a smaller spatial extent of the turbulent/outflowing cold molecular gas compared to the less-dense warm ionised gas (see \citealt{nayakshinZubovas12, mukherjee16, mukherjee18a}). Furthermore, it is interesting to see the asymmetry (i.e., unipolar nature) of the CO(3--2) cones. One of the possible reasons could be that the galaxy disk maybe optically thicker to the CO(3--2) emission, thus hiding the other side of the turbulence, or the molecular gas could itself be intrinsically asymmetric as predicted by simulations at higher redshifts (\citealt{gaborBournaud14}) and observations at lower redshifts (\citealt{lutz20}). 

Our results are in qualitative agreement with recent hydro-dynamical simulations showing that when a radio jet propagates through an inhomogenous medium, its impact can go beyond the radio-jet axis. This happens because as the jets progress, they launch outflows close to them which disturb the local velocity flows and kinematics. However, they also inflate a spherical bubble of jet plasma that can lift the multi-phase gas and cause turbulence, which would naturally be collimated along the minor axis of the host galaxy disk (\citealt{sutherlandbicknell07, wagnerbicknell11, wagner12, mukherjee16, mukherjee18a, talbot21}). This would mean the disturbed gas would be extended beyond the radio emission (also see \citealt{Zovaro19}). \cite{mukherjee18a, mukherjee18b} further predict a stronger impact on the host-galaxy's ISM when low power jets are inclined into the plane of the galaxy disk and have only moderate power, which matches the characteristics of the jets in J1316+1753. Within these simulations, as the jets flood through the inter-cloud galaxy disk, they cross and interact more with the inter-stellar medium due to their lower inclination. We thus attribute the turbulent, high velocity-dispersion gas being vented out perpendicular to the galaxy disk in J1316+1753 to be originating as a consequence of the jets propagating through the ISM. Our observations are less consistent with a model where a quasar-driven shocked wind propagates through the ISM and causes both the radio emission and the turbulence (e.g., \citealt{hwang18}).

\subsection{Feedback effects on the host galaxy}\label{sec: feedback}

While the effect of radio jets is well studied and established at the scales of galaxy halos, particularly for massive early-type galaxies and in cluster environments (\citealt{mcnamara12, hardcastle20}), it is less clear to what extent galaxies can be directly affected by low power jets confined within the ISM. Recent simulations show that, while on one hand density enhancements near the jet axis can make gravitational collapse of clouds more probable, thereby {\em supporting} star-formation; an overall increase in turbulent kinetic energy may also {\em lower} the star-formation rate (\citealt{mandal21}). From an observational perspective, studies have shown how jets could potentially suppress or regulate the star-formation in the host galaxy by introducing turbulence in the ISM or causing multi-phase outflows (e.g., \citealt{nesvadba10, nesvadba11, alatalo15}). On the contrary, studies have also shown a jet-induced positive feedback being driven by the shocks from the jets causing the gas to compress and thus inducing star-formation (e.g., \citealt{bicknell00, salome15, lacy17}). Hence, a jet may cause both a positive and a negative feedback on its host galaxy. Since we observe significant jet-ISM interactions and outflowing material in our observations we now investigate the possible feedback effects arising from these on the host galaxy.

\subsubsection{Impact of the jets on the molecular star-forming gas}\label{sec: energetics}

To understand the impact of the jets in J1316+1753, we first quantify how much energy has been available from the jets over the lifetime of the observed outflows, using order-of-magnitude estimates. We assume a single jet episode, since we lack any evidence of multiple jet episodes. Under the assumption that the turbulent outflowing ionised gas is launched by the jet, we use it's kinematics to estimate the lifetime. The outflow velocity in the ionised gas is estimated to be $V_{\rm out}=800$\,km\,s$^{-1}$, at a distance of $R_{\mathrm{out}}=7.5$\,kpc (see Section \ref{sec: res_gal_ion}).\footnote{We note that if we use the properties of the CO gas instead of the [O~{\sc iii}] we would obtain an estimated lifetime of $t_{\mathrm{out}}$\,$\sim$\,6\,Myr, which is sufficiently close to that derived from [O~{\sc iii}] for our order-of-magnitude estimates} If we then estimate the outflow lifetime as: 
\begin{equation}
t_{\mathrm{out}} \approx R_{\mathrm{out}}/v_{\mathrm{out}},
\end{equation}
we obtain $t_{\mathrm{out}}\sim$9\,Myr. Using the value of the jet power ($P_\mathrm{jet}=10^{44}$\,erg\,s$^{-1}$) estimated in Section \ref{sec: jetprop}, we estimate the total energy from the jet over this period as:
\begin{equation}
E_{\mathrm{jet}} = P_{\mathrm{jet}}\,\times\,t_{\mathrm{out}} \sim3\,\times\,10\,^{58}\,{\rm erg}.
\end{equation}

Next, we estimate the kinetic energy, $E_{\rm mol,gas}$ in the molecular gas phase (traced via CO) of the galaxy disk following \citealt{nesvadba10}, i.e., using the relationship between molecular gas mass, $M_{\rm mol,gas,}$ and velocity-dispersion:
\begin{equation}
    E_\mathrm{mol,gas} = 0.5\,\times\,M_\mathrm{mol,gas}\,\times\,\sigma_\mathrm{mol,gas}^2.
\end{equation}
We focus on the molecular gas phase since the molecular gas reservoirs play a determining role in the AGN-galaxy co-evolution, as it is this gas that is redistributed to promote the star-formation in the galaxy and also for fuelling the black-hole growth in the AGN (\citealt{carilli13,vito14}). To determine the molecular gas mass, we first convert our measured CO fluxes into a CO\,(3--2) emission-line luminosity ($L'_{\rm CO(3-2)}$ [K km s$^{-1}$ pc$^{-2}$]), following \citealt{solomon97} and then into $L'_{\rm CO(1-0)}$, by adopting two different values for the line brightness temperature ratio, r$_{31}$\,=\,0.5 and r$_{31}$=1 (following \citealt{weiss07, riechers11, bolatto13, sargent14}). The former value is more consistent with quiescent star-forming gas, while the latter is for disturbed gas affected by the AGN. Finally, the molecular mass was estimated by using two values of the H$_2$-to-CO conversion factor of $\alpha_{\rm CO}$\,$\approx$\,4 M$_{\odot}$/(K km s$^{-1}$ pc$^{-2}$) and $\alpha_{\rm CO}$\,$\approx$\,0.89 M$_{\odot}$/(K km s$^{-1}$ pc$^{-2}$). The choice of the former $\alpha_{\rm CO}$ is justified in \cite{jarvis2020} and is consistent with that found in the Milky Way and assumed for quiescient gas. The latter choice is for typical high-redshift, high star-forming, quasar host galaxies, following (\citealt{solomon87, solomonVandenBout05, carilli13, bolatto13}). Although uncertain, these conversion factor choices are sufficient for our order-of-magnitude estimates (see further discussion in \citealt{jarvis2020}). If we focus on the pixels inside the 2.5\,kpc region showing the disturbed outflowing molecular gas (i.e., pixels with W$_{80}\,\geq$\,400\,km\,s$^{-1}$, shown in Figure \ref{fig: alma_zoom}), the range of molecular mass in the broad component was estimated to be \,$\sim$\,3.1\,$\times$\,10$^{7}$\,M$_\odot$\,--\,2.8\,$\times$\,10$^{8}$\,M$_\odot$ (using the different conversion factors). Hence, this broad component is seen to consist $\sim$\,1\,\% of the total CO(3--2) flux. Furthermore, using the measured velocity, $\sigma_\mathrm{mol,gas}$\,$=$\,230\,km\,s$^{-1}$, the total kinetic energy was estimated to be E$_\mathrm{mol,gas}\sim\,1.6\,\times\,10^{55}$\,erg\,--\,\,1.4\,$\times\,10^{56}$\,erg. 

Hence, with a jet power higher by two orders of magnitude than the molecular kinetic energy, the jets are sufficient to drive the observed outflow. However, comparing the outflowing molecular gas mass to the total molecular mass in the galaxy, which is estimated as $\sim$\,10$^{10}$\,M$_\odot$ (see \citealt{jarvis2020}), we conclude that only a very small fraction of the total  molecular gas mass is residing in this turbulent, outflowing material. This suggests little in-situ disruption to the overall molecular gas disk (also see \citealt{RamosAlmeida22}).

We further explore the possible fate of the molecular gas by comparing this energy to the galaxy binding energy. We used the stellar mass estimate of M$_{\star}$= 10$^{11}$\,M$_{\odot}$ from \citealt{jarvis2019} and the stellar velocity-dispersion of $\sim$\,300\,km\,s$^{-1}$ (computed as an average value for the central bins in Figure \ref{fig: muse_alma_gist}). Hence, the binding energy of the galaxy is estimated as\footnote{We note that if we estimate the binding energy following the universal relationship between stellar mass and binding energy presented in \cite{shi21} we obtain the same order-of-magnitude result.}, 
\begin{equation}
E_{\star} = M_{\star}\,\times\,\sigma_{\star}^2\,\approx\, 10\,^{59}\,{\rm erg}.    
\end{equation}

Comparing the jet power with the binding energy of the galaxy, we thus conclude that the current jet episode will be limited in its ability to completely unbind the molecular gas from the host galaxy. However, negative AGN feedback may still manifest in several forms (see \citealt{federrathklessen12,costa20,mandal21}), for example: (i) by quickly ejecting the dense gas from only the nuclear regions, as soon as the AGN outburst begins; (ii) by gradually suppressing the halo gas accretion onto the host galaxy; (iii) by inducing turbulence that provide additional support against the collapse, thereby strongly suppressing star-formation. Indeed, recent jet-ISM simulations do not predict a complete unbinding of host galaxy star-forming material, but rather a suppression of star-formation at the global galaxy scales by a factor of a few due to the ablation and fragmentation of the clouds (\citealt{mukherjee18b,mandal21}).

\subsubsection{Evidence for feedback from stellar kinematics}\label{sec: positivefeedback}

We now focus on the possible signatures of feedback on the stellar properties of J1316+1753. We can clearly see a bright elongated bulge-like structure in the stellar continuum emission (Figure \ref{fig: J1316}, left-panel). Following Section \ref{sec: angle_size}, we fitted a two-dimensional gaussian to the surface brightness profile of this bulge-shape, and estimated a position angle of 130\,\degree, which is remarkably close to the position angle of the jet axis, at 120\,\degree\,(red dashed line in Figure \ref{fig: muse_zoom} and \ref{fig: alma_zoom}). In addition, the regions with highest stellar dispersion (i.e., $\sim$290--360\,km\,s$^{-1}$) follow, almost exactly, the jet axis (see the left-bottom panel of Figure \ref{fig: muse_alma_gist} and the galaxy schematic Figure \ref{fig: schematic}). We discuss two possible explanations of these observations.

We have seen clear signatures of jet-ISM interactions. A compression of the ISM gas caused by the jets could therefore trigger episodes of star-formation, i.e. positive feedback. Stars formed during feedback will initially have motions which are significantly perturbed from regular circular stellar kinematics, and hence they may contribute to increasing the stellar velocity-dispersion and could alter the stellar structure of the host galaxy. This could be due to either directly induced positive feedback by jets or outflows, or because of stars that have formed in outflowing/turbulent material (see \citealt{ishibashi13,zubovas13,Dugan14,maolino17,gallagher19}). This may explain the increased velocity disperion along the jet axis from our observations. Furthermore, at the jet termini, we can see gas velocities of $-$235\,$\pm$\,1.3 and 206\,$\pm$\,15 km\,s$^{-1}$, which entails to the jets pushing the gas with this velocity (the velocities seen in the two regions of enhanced [O~{\sc iii}] SNR, or regions of jet-ISM interactions in Figure \ref{fig: muse_zoom}). Interestingly, this roughly matches the higher stellar velocity-dispersion seen in these regions ($-$268\,$\pm$\,18 and 240\,$\pm$\,14\,km\,s$^{-1}$, see Figure \ref{fig: muse_alma_gist}). Detailed analyses of the stellar populations found in the different stellar kinematic components would be required to further test this scenario.

Another possible reason for this high stellar dispersion could be due to the sudden removal of gas in the very central regions, thereby causing a decrease in the local gravitational potential that may lead to ballistic orbits of central stars (see \citealt{vlugtCosta19}), thereby causing an apparent increase in the stellar velocity-dispersion. However, it is not immediately clear why this would cause enhanced stellar velocity-dispersion along the jet axis and detailed dynamical modelling (accounting for the relative mass in stars, dark matter and gas) would be required to test this scenario further.

\subsection{Implications of our results for `radio-quiet' AGN feedback}\label{sec: lit_comp}

\begin{figure}
	\includegraphics[width=\columnwidth]{ 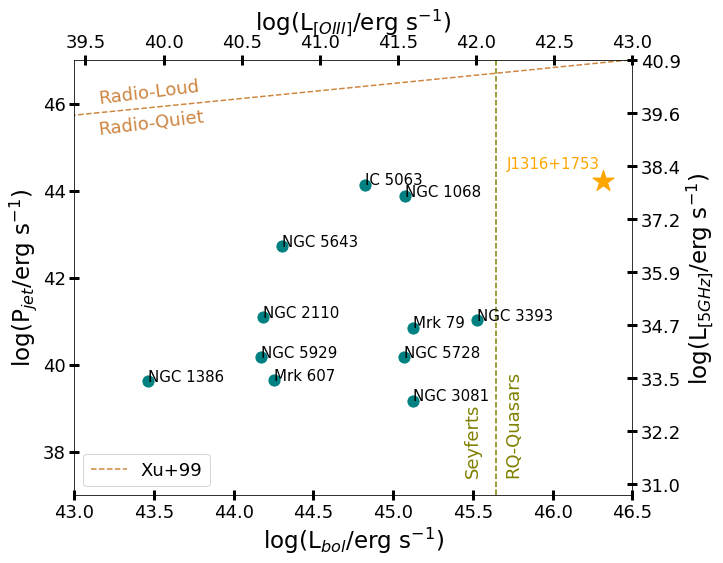}
    \caption{A comparison of jet power ($P_{\mathrm{jet}}$; scaled directly from 5\,GHz radio luminosity) versus $L_{\rm Bol}$ for radio-quiet AGN-host galaxies from literature, which show high velocity-dispersion gas, perpendicular to radio jets (see Section \ref{sec: lit_comp}). Although this effect has been seen numerous times in the Seyfert-luminosity regime, J1316+1753 target clearly lies in the quasar regime as identified using our definition ($L_{\rm [O III]}\geq$10$^{42.1}$ erg s$^{-1}$, dashed line). The dashed orange line traces the division between `radio-quiet' and `radio-loud' AGN from \citealt{xu1999}.}
    \label{fig: overview}
\end{figure}

Jet-ISM interaction similar to the one reported here have been identified in  several other radio-quiet AGN host galaxies. We compare our source to a list of 13 targets, compiled by \cite{venturi2021}, which show a similar observation of high velocity-dispersion gas in the perpendicular direction (Section~\ref{sec: jet_ism}). In Figure \ref{fig: overview}, we demonstrate this comparison of J1316+1753 in the $L_{\rm bol}$\,-\,$P_{\mathrm{jet}}$ plane. We derived the $P_{\mathrm{jet}}$ from the 5\,GHz radio luminosities\footnote{For three of the sources where 5\,GHz values were not available, we used 1.4\,GHz values and converted them using a simple power law and an average spectral index \(\alpha=-0.7\) to calculate 5\,GHz luminosities. References used for radio luminosities: \citealt{jarvis2019,litradio1,litradio2,litradio3,litradio4,litradio5,litradio6,litradio7}} in literature, following Equation \ref{eq: pjet}. This approach is sufficient for our order-of-magnitude investigation of the parameter space in Figure \ref{fig: overview}. We used the $L_{\mathrm{Bol}}$/$L_{\mathrm{[OIII]}}$\,=\,3500 ratio following \cite{heckman04} to obtain the $L_{\mathrm{Bol}}$ values from the $L_{\mathrm{[OIII]}}$ values in literature\footnote{We note that since we could not obtain the $L_{\mathrm{[OIII]}}$ values for two of the sources, they are excluded from Figure \ref{fig: overview}. This does not affect our results since the two excluded sources have X-ray flux emission that might be rather associated to X-ray binaries than with accretion on a SMBH. References used for $L_{\mathrm{[OIII]}}$: \citealt{jarvis2019,mm07,sc03,wh92}.}. 

It can be clearly seen that the target we study here has a higher bolometric luminosity than those typically identified to have these type of jet-ISM interactions.  Although J1316+1753 is at the upper end of jet powers, with P$_{\rm jet}$=10$^{44}$\,erg\,s$^{-1}$ compared to the comparison sample in Figure~\ref{fig: overview} (which reach P$_{\rm jet}$<10$^{44}$\,erg\,s$^{-1}$), our target still lies in the radio-quiet regime. However, in contrast to the other sources, this target is a quasar, with $L_{\rm [O III]}$\,$\ge
$\,10$^{42.1}$ erg\,s$^{-1}$\,(shown as dashed vertical line in the figure). Therefore, we would like to stress that for quasars and in general if $L_{\mathrm{Bol}}$\,$>>$\,$P_{\mathrm{jet}}$, jets could still be present as strong sources of feedback and hence should not be ignored (also see \citealt{VillarMartin17}). Indeed with the increasing availability of deep, high resolution imaging, an increasing number of systems are also showing the presence of jets (e.g., \citealt{jarvis2019, pierce20, villarmartin21}).

In summary, we have shown that for J1316+1753, low power jets are the dominant mechanism for the central AGN to have an impact on the host galaxy's ISM, even when the system is bolometrically powerful. Interestingly, by comparing the hydrodynamical simulations of individual galaxies that invoke a jet (\citealt{mukherjee18b,mandal21}) with those that invoke a quasar-wind (\citealt{costa14,costa20}), suggests that both mechanisms may have a very similar impact on the host galaxy (also see \citealt{fauchergiguere12, talbot21}); specifically, both localised enhanced star-formation (due to compression of gas) and more global, gradual suppression of star-formation. 
Understanding which of these modes dominate for quasar host galaxies, and how representative J1316+1753 is across the quasar population, will require further studies combining high quality radio imaging with spatially-resolved measurements of the multiphase gas. 
We plan such a study across the wider QFeedS sample, where it will become possible to investigate trends with jet power and jet inclination angle, which are two crucial parameters according to simulations (\citealt{mukherjee18b}). Further careful studies of the role of {\em low power jets} for AGN feedback is important because lower power jets are very common, compared to their more powerful counterparts, in high mass galaxies (including quasars) even if it is harder to decouple their radio emission from that produced by star-formation  (\citealt{Guerkan19,Sabater19,Macfarlane21}).

\section{Conclusions}\label{sec: conclusions}

We present MUSE and ALMA data of J1316+1753, a luminous $z$\,=\,0.15, type-2 quasar selected from the Quasar Feedback Survey \citep[][]{jarvis21}. This target represents one of the most [O~{\sc iii}] luminous sources from the survey ($L_{\rm [O III]}$=10$^{42.8}$\,erg\,s$^{-1}$) and exhibits a broad [O~{\sc iii}] emission-line profile (FWHM$\sim$1300\,km\,s$^{-1}$; Figure~\ref{fig: mullaneyplot}; Figure~\ref{fig: fits}). Radio imaging of this source reveals low-power radio jets ($P_{\rm jet}\sim$10$^{44}$\,erg\,s$^{-1}$) that  are compact, reaching only 1\,kpc in projected distance from the core. Furthermore, the jets are inclined into the plane of the host galaxy disk (see Section \ref{sec: jetprop} and Figure~\ref{fig: J1316}). 

Our data enables us to map the stellar kinematics (traced with stellar absorption features), warm ionised gas properties (traced with optical emission-lines) and the cold molecular gas properties (traced with the CO(3--2) emission-line; see Figure~\ref{fig: fits}).  On galaxy scales, both the molecular gas and ionised gas broadly follow the stellar gravitational motions (Figure~\ref{fig: muse_alma_gist}). However, across the central few kiloparsecs, both gas phases reveal high velocity non-gravitational motions and we observe evidence of jet-induced feedback. Specifically:

\begin{itemize}
    \item[1.] \textbf{Jet-ISM interactions}\\
We observe two bright and high velocity offset ionised gas components (separated by 441 km\,s$^{-1}$) concentrated at the positions of the jet hot spots and that appear to propagate away from the jets, along the jet axis (see Figure~\ref{fig: muse_zoom}). Furthermore, a $-100$\,km\,s$^{-1}$ change in the molecular gas velocity is observed just beyond the brighter radio hot spot, with tentative evidence for depleted CO\,(3--2) emitting gas at the same location (see Figure~\ref{fig: alma_zoom}).    \\
    
    \item[2.] \textbf{Multi-phase turbulent gas, driven perpendicular to the galaxy}\\
Ionised gas with very high velocity-dispersion (i.e., $W_{80}$\,=\,1000\,--\,1300\,km\,s$^{-1}$) is seen to propagate outwards in a bi-cone from just behind the radio hot spots, travelling perpendicular to the galaxy disk and seen in projection as extending to at least 7.5\,kpc from the nucleus (see Figure~\ref{fig: muse_alma_gist} and ~\ref{fig: muse_zoom}). The highest inferred electron densities of the ionised gas are found within these bi-cones (as inferred from the [S~{\sc ii}] doublet; Figure~\ref{fig: muse_zoom}).  This turbulent gas is also seen in the molecular gas phase. However, it is 3 times less extended (and only in one direction) with 3 times lower velocity-dispersion (i.e., W$_{80}\sim$\,400\,km\,s$^{-1}$) compared to ionised gas phase (see Figure~\ref{fig: muse_alma_gist} and \ref{fig: alma_zoom}). \\
    
    \item[3.] \textbf{Strong spatial connection between the jets and stellar properties}\\
 We see a close alignment of the position angle of the stellar bulge with the radio-jet axis ($<$\,5\degree; see Figure \ref{fig: J1316}, left-panel). Furthermore, the regions with the highest stellar velocity-dispersion  (i.e., $\sigma_{\star}$ are seen to be lagging behind the jets, following the jet axis (see Figure \ref{fig: muse_alma_gist}, bottom-left panel).
\end{itemize}

Our observations provide strong evidence for low power radio jets, inclined into the galaxy disk, having a direct impact on the multi-phase ISM inside the host galaxy of this type-2 quasar. Our observations are qualitatively consistent with the simultaneous positive and negative feedback effects observed in hydrodynamics simulations of jet-ISM interactions \citep[][]{mukherjee18a, mukherjee18b, talbot21, mandal21}. Specifically: (1) as the inclined, lower power jets move through the galaxy, they strongly interact with the ISM; (2) highly turbulent material is stripped and escapes above and below the galaxy disk, removing gas from the host galaxy and; (3) as the jets propagate, they compress the gas in the disk also causing new stars to form in these regions, contributing to the formation of the stellar bulge (see Section \ref{sec: positivefeedback}). Although we only have indirect evidence, this final factor may be the cause of the high velocity-dispersion stars located behind the jets we have observed. 

Interestingly, the same qualitative behaviour of outflows and positive feedback in the inner few kiloparsecs of galaxies is predicted by simulations invoking quasar-driven disk winds (\citealt{costa20}). Whilst jets have been seen to be dominant in lower power AGN (See Figure~\ref{fig: overview} and Section~\ref{sec: lit_comp}), our observations reveal that jets can be the dominant feedback mechanism, even for this bolometrically luminous source, i.e., for an AGN currently in a `quasar mode'. To understand the relative role and impact of jets and winds in all quasars requires a similar multi-wavelength study of the wider population.


\section*{Acknowledgements}
The authors would like to thank the referee for their comments which helped improved the quality of the manuscript. PK and SS acknowledge the support of the Department of Atomic Energy, Government of India, under the project 12-R\&D-TFR-5.02-0700. DMA and ACE acknowledge support from STFC (ST/T000244/1). We would like to acknowledge the valuable insights from Dimitri Gadotti in helping us with the stellar spectral fits. This research made use of Astropy,\footnote{http://www.astropy.org} a community-developed core Python package for Astronomy \citep{astropy2013, astropy2018}.  

This paper makes use of the following ALMA data: ADS/JAO.ALMA\#2018.1.01767.S. ALMA is a partnership of ESO (representing its member states), NSF (USA) and NINS (Japan), together with NRC (Canada), MOST and ASIAA (Taiwan), and KASI (Republic of Korea), in cooperation with the Republic of Chile. The Joint ALMA Observatory is operated by ESO, AUI/NRAO and NAOJ.

\section*{Data Availability}
The MUSE and ALMA data presented in this analysis were accessed from the ESO archive (\url{http://archive.eso.org/scienceportal/home}) under the proposal ids: 0103.B-0071 and 2018.1.01767.S, respectively. The VLA image used in this work is available at Newcastle University's data repository (\url{https://data.ncl.ac.uk}) and can also be accessed through our \href{https://blogs.ncl.ac.uk/quasarfeedbacksurvey/data/}{Quasar Feedback Survey website}.



\bibliographystyle{mnras}
\bibliography{citations_list} 


\bsp	
\label{lastpage}

\appendix

\section{Voronoi Binning Map}\label{app: voronoi}
\begin{figure}
\centering
\includegraphics[width=\columnwidth]{ 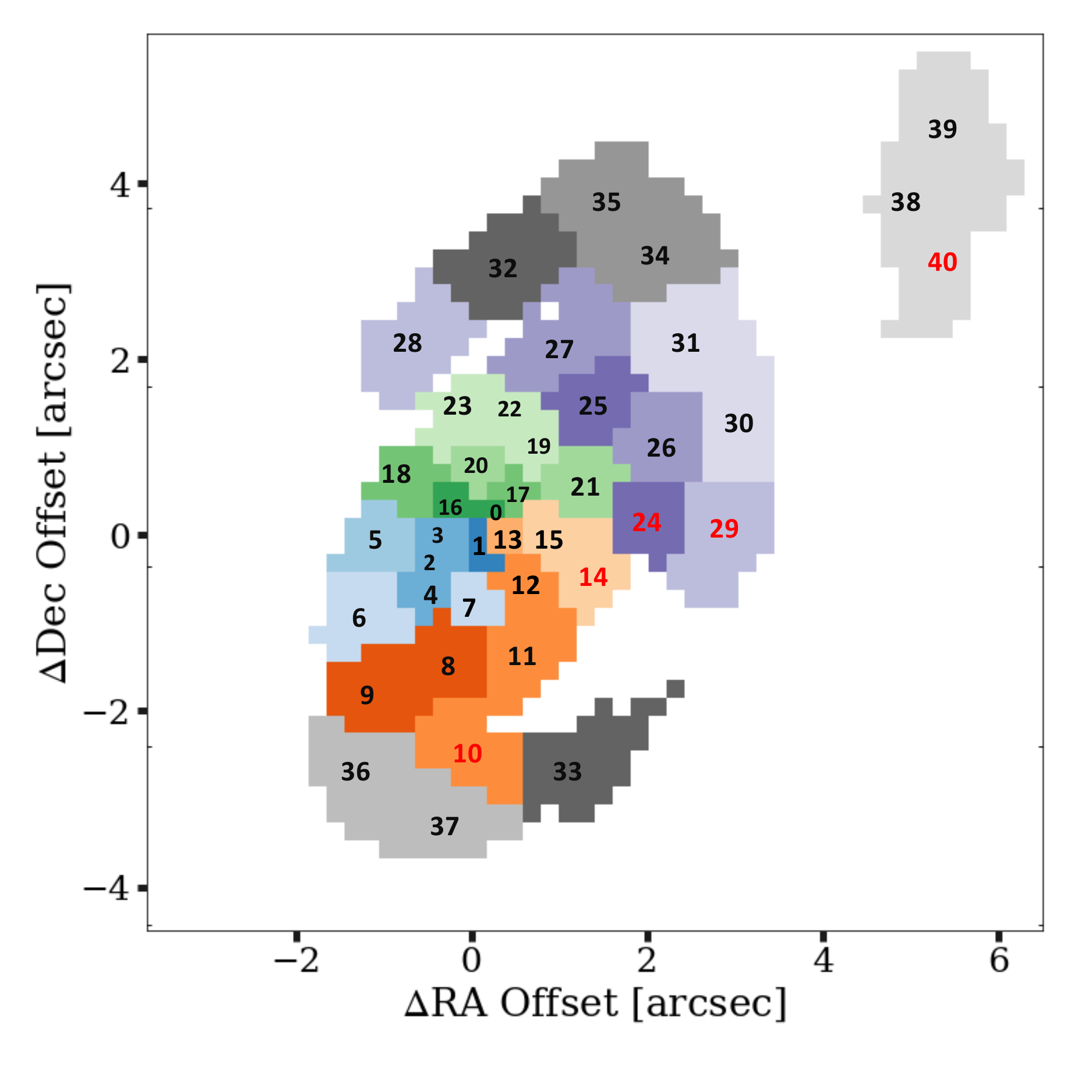}
\caption{A map of the 41 Voronoi Bins obtained using a SNR of 5, as described in the Section \ref{sec: gist}. The same Voronoi mapping is used for the analysis of the MUSE and ALMA data at the galaxy-scale, as summarised in Section \ref{sec: res_gal_ion} and \ref{sec: res_gal_mol}, respectively. Five of the bins (no. 10, 14, 24, 29, 40) that were excluded from the analysis of the molecular gas due to low SNR are labelled in red. Also it can be seen here that the bins 38, 39, 40 correspond to the companion galaxy while the rest belong to the main galaxy as labelled in Figure~\ref{fig: J1316}.}
\label{fig: voronoi_map}
\end{figure}

In Figure \ref{fig: voronoi_map}, we show a map with the spatial distribution and numbering system of the Voronoi bins, produced from the Voronoi tesselation routine, as explained in Section \ref{sec: gist}. 

\section{Electron Densities}\label{app: ne}
\begin{figure*}
	\includegraphics[width=0.9\textwidth]{ 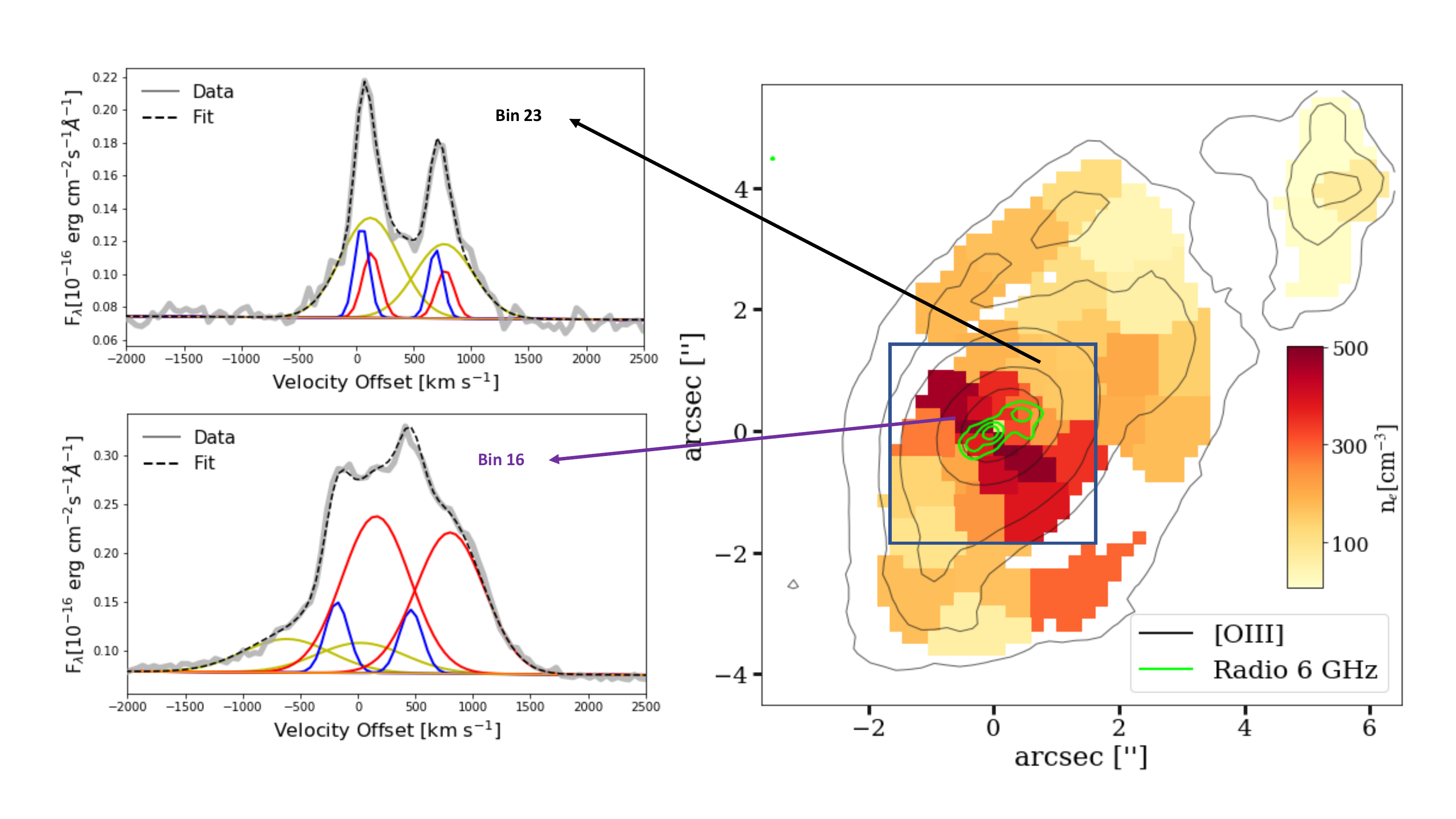}
    \caption{A representation of the spatial distribution of electron densities ($n_e$) using the [S~{\sc ii}] doublet method. {\em Right panel:} A map colour-coded by $n_e$, estimated for each of the Voronoi bins, mapped over the entire galaxy. The contours are same as described in Figure \ref{fig: J1316}. The small inset box shows the central-scale, for which the map is presented in Figure \ref{fig: muse_zoom}. {\em Left panels:} Example [S~{\sc ii}]-emission-line profiles and their fits are shown for regions of low (Bin 23) and high (Bin 16) electron densities. The data and fitting-components are displayed using the same convention as for Figure \ref{fig: fits}.}
    \label{fig: ne}
\end{figure*}

We estimated the spatial variations in electron density by using the ratio of the  [S~{\sc ii}]$\lambda\lambda$6716,6731 emission-line doublet, assuming a temperature of $T_e$\,=\,10$^4$\,K (\citealt{osterbrockfernland2006}). As described in Section \ref{sec: emline}, we fit the [S~{\sc ii}] doublet with Gaussian components: three components for each line, as shown in the left-panel in Figure \ref{fig: ne} for two example bins. We note that the [S~{\sc ii}] emission line fits were fit independently for their flux, velocity and linewidths, without any influence of the [O~{\sc iii}] parameters obtained in Section \ref{sec: emline}. The right-panel of Figure \ref{fig: ne} shows a map of the estimated electron density values on the galaxy-scale using the Voronoi bins that were defined on the stellar continuum in Section~\ref{sec: gist}. We further perform this analysis on the central-scale using individual MUSE spaxels, the results of which are shown in the bottom-right panel of  Figure \ref{fig: muse_zoom}. We note that the [S~{\sc ii}] doublet is only sensitive to electron densities in the range 50\,$\lesssim$n\,$_e$\,$\lesssim$\,5000\,cm$^{-3}$ (\citealt{osterbrockfernland2006}). Therefore beyond this range, we indicate the electron densities are unconstrained in these regions through grey pixels in Figure \ref{fig: muse_zoom} and we discuss this limitation further below.

Interestingly, both the central-scale (10\,kpc) and the galaxy-scale analyses (26\,kpc) reveals that higher electron densities (i.e., 
$\sim$500--600cm$^{-3}$) are observed  perpendicular to the radio jets. Furthermore, the regions of the highest inferred electron densities are remarkably co-spatial with the most disturbed ionised gas (see Section \ref{sec: res_nuc_ion}), whilst the spaxels outside of this cone exhibit more modest electron densities (i.e., $\sim$\,150 cm$^{-3}$). This has also been observed by other studies (e.g.; see \citealt{fluetsch21,davies20, mingozzi19}). However, we urge caution in interpreting the absolute values of the electron densities inferred using this approach. For example, the [S~{\sc ii}] doublet is limited in its diagnostic power for tracing electron densities and may give results that are about 1-2 orders lower in magnitude than the electron densities derived using trans-auroral [O~{\sc ii}] and [S~{\sc ii}] lines (for details, see \citealt{harrison18}). A recent study by \citealt{davies20} also found significantly lower electron densities using this method as compared to the auroral and trans-auroral lines (also see, \citealt{rose18, harrison18, shimizu2019, baronNetzer19}). 


\section*{Supporting Information}\label{suppinfo}
Supplementary data are presented in a companion PDF document. \\
\textbf{Appendix A}: The following spectral fits are shown for all Voronoi Bins:
\begin{itemize}
    \item Ionised gas fits: [O~{\sc iii}];
    \item Molecular gas fits: CO\,(3--2);
    \item Stellar continuum fits.
\end{itemize}


\end{document}